\documentclass[]{acta}
\usepackage{supertabular,lscape,epsfig}
\usepackage{amssymb}
\usepackage{amsmath}
\usepackage{graphicx}
\usepackage{color}

\SetPages{0}{0}

\SetVol{00}{2024}

\begin{document}

\setcounter{page}{1}

\begin{Titlepage}
\Title{Phosphorus Abundances of B-Type Stars in the Solar Neighborhood}
\Author{Yoichi~~T~a~k~e~d~a}{11-2 Enomachi, Naka-ku, Hiroshima-shi, Japan 730-0851\\
e-mail: ytakeda@js2.so-net.ne.jp}

\Received{Month Day, Year}
\end{Titlepage}

\Abstract{
Phosphorus abundances of $\sim 80$ apparently bright sharp-lined early-to-late B-type 
stars on the upper main sequence are determined by applying the non-LTE analysis 
to the P~{\sc ii} line at 6043.084~\AA, with an aim of getting information on the P 
abundance of the galactic gas (from which these young stars were formed) in 
comparison with the reference solar abundance ($A_{\odot} \simeq 5.45$).
These sample stars turned out to be divided into two distinct groups with respect 
to their P abundances: (1) chemically peculiar late B-type stars of HgMn group show 
considerable overabundances of P (supersolar by $\sim$~0.5--1.5~dex), the extent of which
progressively increases with $T_{\rm eff}$. (2) In contrast, the P abundances of 
normal B-type stars are comparatively homogeneous, though a notable difference is 
observed between the LTE and non-LTE cases. Although their LTE abundances are near-solar,
a slight gradual trend with $T_{\rm eff}$ is observed. However, after applying the negative 
non-LTE corrections (amounting $\sim$~0.1--0.5~dex), this $T_{\rm eff}$-dependence is 
successfully removed, but the resulting non-LTE abundances (their mean is $\simeq 5.20$) 
are appreciably underabundant relative to the Sun by $\sim$~0.2--0.3~dex. 
The cause of this systematic discrepancy (contradicting
the galactic chemical evolution) is yet to be investigated.
}
{Galaxy: solar neighborhood -- stars: abundances -- stars: chemically peculiar -- 
stars: early-type -- stars: population I}

\section{Introduction}

Astrophysical interest in the cosmic abundance of phosphorus (P; $Z = 15$) is rapidly 
growing these days. While one reason is that the mechanism of how this element is synthesized
in the chemical evolution of the Galaxy is not yet well understood, another intriguing    
motivation lies in its astrobiological context.
That is, P is (along with H, C, N, O) an indispensable key element 
for life, which is the backbone of nuclear acids (RNA, DNA) or cell 
membranes, and plays a significant role in producing/reserving 
the vital energy via ATP (Adenosine TriPhosphate). 
Especially, the original P abundance of protoplanetary material (or star-forming gas)  
in comparison with that of the Sun (comparatively P-rich) is an important factor, 
because substantially subsolar case would make it difficult to leave sufficient amount P 
for life on the surface of planets due to its strongly partitioning nature in the planetary 
core (Hinkel et al. 2020).

Accordingly, not a few stellar spectroscopists devoted their energies to P abundance determinations
since 2010s, and more than a dozen papers have been published during the short period of 
the past decade (see, e.g., Table~1 of  Sadakane \& Nishimura 2022, Sect.~1 of Maas et al. 2022, 
and the references therein). These authors established the phosphorus abundances of
various late-type (FGK-type) stars by using neutral P~{\sc i} lines in the near-infrared 
region ($Y$-band or $H$-band) or in the ultraviolet region ($\sim 2135$\AA). 

It has thus revealed that [P/Fe] (logarithmic P-to-Fe ratio) tends 
to progressively increase with a decrease in the metallicity ([Fe/H]) 
like $\alpha$-elements for stars of disk population (from [P/Fe]~$\sim 0$ 
at [Fe/H]~$\sim 0$ to [P/Fe]~$\sim +0.5$ at [Fe/H]$\sim -1$)  
whereas it turns to drop again in the metallicity regime of halo 
population ([P/Fe]~$\sim 0$ at [Fe/H]~$\lesssim -2$), as shown 
in Fig.~1 of  Bekki \& Tsujimoto (2024).
While several theoreticians tried to explain this trend of [P/Fe] 
mainly in terms of the P production by core-collapse supernovae,
Bekki \& Tsujimoto (2024) recently argued that oxygen--neon (ONe) 
novae (triggered in close binary systems including heavier white 
dwarfs) should play a significant role in the galactic nucleosynthesis 
of P. 

However, these studies are directed only to comparatively cool stars of lower mass (typically 
around $\sim 1 M_{\odot}$) which have ages on the order of $\sim 10^{9}$--$10^{10}$~yr.
Meanwhile, much less efforts have been made to young hotter stars, such as B-type stars 
of $\sim$~3--10~$M_{\odot}$ (reflecting the gas composition of the Galaxy in the more 
recent past; i.e. several times $\sim 10^{7}$--$10^{8}$~yr ago). Actually, phosphorus 
abundance determinations of B stars are generally scarce, excepting HgMn stars (chemically 
peculiar late B-type stars showing considerable overabundance of P by up to $\sim$~1--2~dex),
for which P abundances have been reported for quite a number of stars as compiled by 
Ghazaryan \& Alecian (2016).  

This is presumably due to the difficulty in finding useful P lines of sufficient
strengths. In the atmosphere of B-type stars, P atoms are mainly in the ionization stages 
of P~{\sc ii} (late--mid B) or P~{\sc iii} (early B), as illustrated in Fig.~1. 
Since available P~{\sc ii} or P~{\sc iii} lines in the optical wavelength 
regions are all of high-excitations ($\chi_{\rm low} > 10$~eV), they are
fairly weak in strength for the case of usual (near-solar) P abundances and thus 
not easy to detect.\footnote{Although strong low-excitation P~{\sc ii} or 
P~{\sc iii} lines do exist in the ultraviolet region, they are not suitable for 
reliable abundance determinations (i.e., too strong and apt to suffer blending).}
 
%Fig. 1
\begin{figure}[h]
\begin{minipage}{150mm}
\begin{center}
\includegraphics[width=9.0cm]{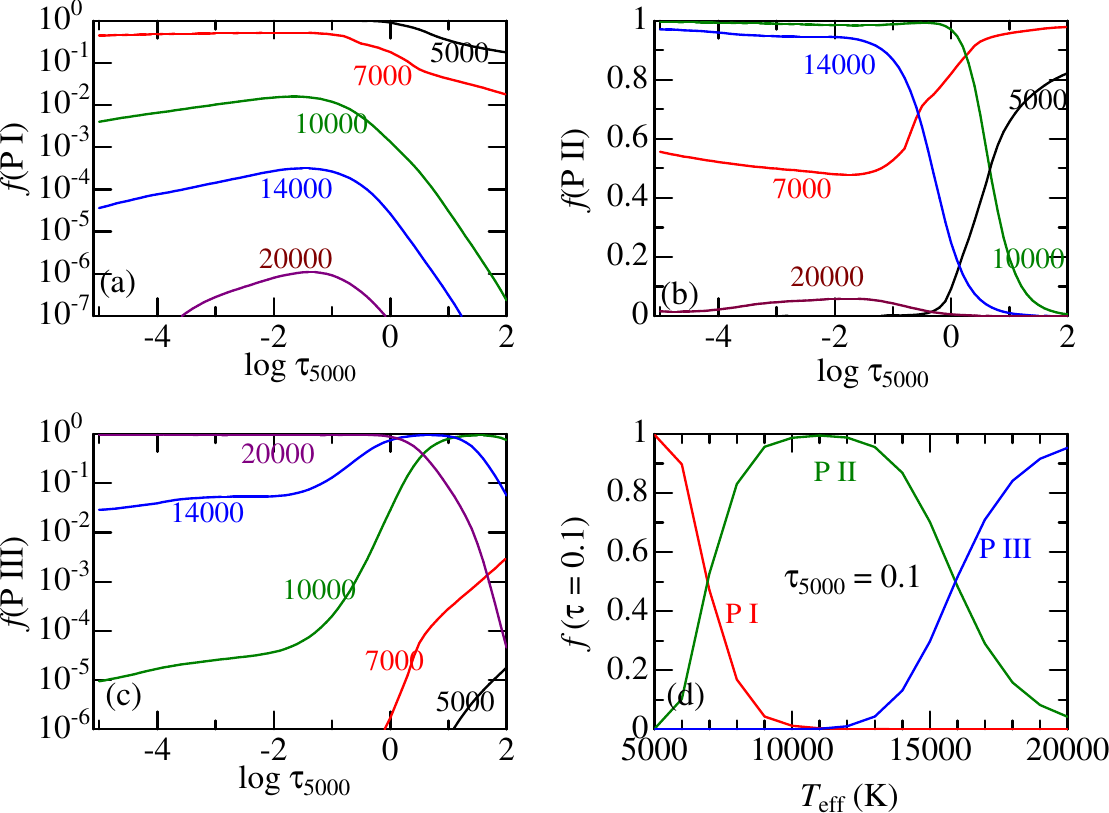}
\end{center}
\FigCap{
Number population fraction ($f$) of (a) neutral, (b) once-ionized, 
and (c) twice-ionized phosphorus species relative to the total P atoms
[e.g., $f$(P~{\sc i}) $\equiv N$(P~{\sc i})/$N_{\rm total}^{\rm P}$],
plotted against the continuum optical depth at 5000~\AA. 
Calculations were done for five $\log g = 4.0$ solar-metallicity models 
of different $T_{\rm eff}$ (5000, 7000, 10000, 14000, and 20000~K) 
as indicated in each panel. The runs of $f$ (at $\tau_{5000} = 0.1$)
with $T_{\rm eff}$ for these three stages are also depicted in panel (d). 
All these calculations were done in LTE (use of Saha's equation).
}
\end{minipage}
\end{figure}

As such, published studies of P abundances for ``normal'' B-type stars are quite limited:
\begin{itemize}
\item
Phosphorus abundances are reported for the well-studied benchmark sharp-lined star 
$\iota$~Her (B3 IV), as summarized in Table~1 of Golriz \& Landstreet 
(2017). However, the results by three studies (Pintado \& Adelman 1993; Peters \& Polidan 
1985; Peters \& Aller 1970; based either on P~{\sc ii} or P~{\sc iii} lines) show 
rather large diversities in $A$\footnote{$A$ is the logarithmic number abundance 
of the element (P) relative to that of hydrogen with the usual normalization of 
$A_{\rm H} = 12$ for H.} from $\sim 5.8$ to $\sim 6.4$; i.e., apparently P-rich
if simply compared with the solar abundance ($A_{\odot} = 5.45$).\footnote{
In this article, Anders \& Grevesse's (1989) solar photospheric P abundance of 
$A_{\odot} = 5.45$ is adopted as the reference, in order to keep consistency with 
Kurucz's (1993) ATLAS9/WIDTH9 program. See Appendix~A for a more detailed discussion
on this subject.}
\item
Pintado \& Adelman (1993) also determined the P abundance of $\gamma$~Peg 
(B2 IV) from P~{\sc iii} lines to be $A \sim 5.4$ (i.e., almost solar). 
\item
In Allen's (1998) abundance studies on early A and late B stars, attempts 
of P abundance determinations for 7 normal stars based on P~{\sc ii} lines in 
the optical region turned out to be unsuccessful, though 5 HgMn stars 
were confirmed to be significantly P-rich ($A \sim$~5.9--7.8). 
\item
Fossati et al.'s (2009) analysis on 2 normal late B-type stars (21~Peg and 
$\pi$~Cet) based on P~{\sc ii} lines yielded [P/H]\footnote{As usual, [X/H] 
is the differential abundance of element X relative to the Sun; 
i.e., [X/H]~$\equiv A_{\rm star}({\rm X})- A_{\odot}({\rm X})$.}~$\sim +0.3$~dex 
(moderately overabundant) for both.
\item
In Niemczura et al.'s (2009) abundance study on late B-type stars (including HgMn 
stars), the P abundances of 3 normal stars were determined as $A$ = 6.01 (HD~49481), 
5.60 (HD~50251), and 5.28 (HD~182198); i.e., nearly solar or moderately supersolar.
\item
Przybilla et al. (2006) derived a near-solar P abundance of $A = 5.53 (\pm 0.06)$ 
for the B-type supergiant $\beta$~Ori (B8 Iae) based on 4 P~{\sc ii} lines.
\end{itemize}

Therefore, phosphorus abundance problem of normal B-type stars is far from being 
settled (near-solar? or somewhat supersolar?), given such insufficient data 
by different investigators for only a small number of stars. What we require 
is a comprehensive study based on a large sample of stars, by which wealthy 
and homogeneous data of P abundances would be accomplished.  

Motivated by this consideration, I decided to conduct an extensive analysis of 
P abundances for $\sim 80$ young B-type stars (including HgMn stars) by using 
the P~{\sc ii} line at 6043.08~\AA\ (hereinafter often 
referred to as P~{\sc ii} 6043; it is the strongest P~{\sc ii} line in 
the optical region and almost free from any appreciable blending), 
in order to estimate the P composition of the galactic gas 
at the time of their formation.

Besides, an emphasis was placed on incorporating the non-LTE affect in 
P abundance determinations based on statistical equilibrium calculations. 
Since non-LTE calculation for P has never been carried out 
so far (to the author's knowledge) and all the past investigations mentioned
above were done based on the assumption of LTE, it would be interesting   
to see how the new non-LTE results compare with the previous ones.

In addition, the significance of the non-LTE effect in the determination of 
phosphorus abundance in the Sun (and FGK-type stars) based on P~{\sc i} lines
in the near-infrared region was also examined as a related topic.
This supplementary analysis is separately presented in Appendix~A. 

\section{Observational Data}

The observational materials used in this study are the high-dispersion 
($R \sim 70000$) and high-S/N ($\sim$~200--700) spectra obtained by HIgh Dispersion 
Echelle Spectrograph (HIDES) placed at the coud\'{e} focus of the 188~cm reflector 
at Okayama Astrophysical Observatory, which are the same as already employed in the 
previous two papers of the author. (i) 64 early-to-late B-type stars (mostly normal stars
but some are late-B chemically peculiar stars) observed in 2006 October, which were 
studied by Takeda et al. (2010; hereinafter referred to as Paper~I) for their O and 
Ne abundance determinations by using O~{\sc i} 6156--8 and Ne~{\sc i} 6143 lines. 
(ii) 21 late B-type stars (HgMn-type peculiar stars and normal stars) observed
in 2012 May, which were analyzed by Takeda et al. (2014; hereinafter Paper~II)
for their Na abundance determinations based on Na~{\sc i} 5890/5896 lines.
See these original papers for more details about the adopted spectra.
The program stars (85 in total) are all sufficiently sharp-lined (projected 
rotational velocities are $v_{\rm e}\sin i \lesssim 60$~km~s$^{-1}$) and are located  
in the solar neighborhood (within $\lesssim 1$~kpc). Their fundamental stellar data 
are listed in Table~1, where two groups (i) and (ii) are presented separately.  

These 85 targets are plotted on the $\log L$ vs. $\log T_{\rm eff}$ diagram in Fig.~2, 
where Lejeune \& Schaerer's (2001) standard theoretical evolutionary tracks 
(solar metallicity, non-rotating models with mass loss; see Sect.~2 therein 
for the details of the input physics they adopted) 
corresponding to different stellar masses are also depicted. 
This figure indicates that the sample stars are in 
the mass range of $2.5 M_{\odot} \lesssim M \lesssim 9 M_{\odot}$.

%Table~1
\setcounter{table}{0}
\begin{table}[h]
\caption{Program stars and the results of the analysis.}
%\small
\scriptsize
\begin{center}
\begin{tabular}{ccccccrrcccc}\hline\hline
\hline
HD\#  & HR\#  & star name  &  Sp.Type  & $^{*}T_{\rm eff}$  & $^{*}\log g$ & $^{\dagger}v_{\rm e}\sin i$ & $W_{6043}$ 
& $A^{\rm L}$   & $\Delta$  & $A^{\rm N}$ & $^{\#}$S/N \\
(1) & (2) & (3) & (4) & (5) & (6) & (7) & (8) & (9) & (10) & (11) & (12) \\
\hline
\multicolumn{12}{c}{[2006 October observations (early-to-late B-type stars)]}\\
 029248 & 1463 & $\nu$  Eri   & B2III      & 22651 & 3.58 &  46 &  $\cdots$ & $\cdots$ & $\cdots$ & $\cdots$ & 450 \\
 000886 & 0039 & $\gamma$ Peg & B2IV       & 21667 & 3.83 &   9 &   1.3 &  5.32 & $-0.04$ &  5.28 & 730 \\
 035708 & 1810 &114 Tau       & B2.5IV     & 21082 & 4.09 &  26 &   3.9 &  5.68 & $-0.25$ &  5.43 & 470 \\
 035039 & 1765 & 22 Ori       & B2IV-V     & 20059 & 3.69 &  10 &   2.2 &  5.34 & $-0.14$ &  5.20 & 460 \\
 042690 & 2205 &              & B2V        & 19299 & 3.81 &  12 &   3.3 &  5.40 & $-0.22$ &  5.18 & 470 \\
 032249 & 1617 & $\psi$ Eri   & B3V        & 18890 & 4.13 &  40 &   3.7 &  5.38 & $-0.30$ &  5.07 & 500 \\
 034447 & 1731 &              & B3IV       & 18480 & 4.10 &   9 &   6.4 &  5.58 & $-0.20$ &  5.38 & 300 \\
 196035 & 7862 &              & B3IV       & 17499 & 4.36 &  35 &   5.6 &  5.45 & $-0.25$ &  5.20 & 300 \\
 043157 & 2224 &              & B5V        & 17486 & 4.12 &  37 &   4.6 &  5.33 & $-0.32$ &  5.01 & 210 \\
 160762 & 6588 & $\iota$ Her  & B3IV       & 17440 & 3.91 &   7 &   4.7 &  5.33 & $-0.30$ &  5.03 & 380 \\
 223229 & 9011 &              & B3IV       & 17327 & 4.20 &  31 &   3.9 &  5.25 & $-0.40$ &  4.86 & 390 \\
 176502 & 7179 &              & B3V        & 16821 & 3.89 &   8 &   7.8 &  5.52 & $-0.24$ &  5.28 & 340 \\
 041753 & 2159 & $\nu$ Ori    & B3V        & 16761 & 3.90 &  28 &   2.7 &  5.02 & $-0.55$ &  4.47 & 520 \\
 025558 & 1253 & 40 Tau       & B3V        & 16707 & 4.29 &  41 &   7.0 &  5.51 & $-0.25$ &  5.27 & 360 \\
 044700 & 2292 &              & B3V        & 16551 & 4.21 &   5 &   6.0 &  5.42 & $-0.30$ &  5.12 & 360 \\
 186660 & 7516 &              & B3III      & 16494 & 3.57 &   9 &   6.5 &  5.38 & $-0.28$ &  5.11 & 370 \\
 181858 & 7347 &              & B3IVp      & 16384 & 4.19 &  17 &   8.5 &  5.58 & $-0.23$ &  5.36 & 340 \\
 023793 & 1174 & 30 Tau       & B3V+F5V    & 16264 & 4.15 &  46 &   8.7 &  5.59 & $-0.23$ &  5.36 & 510 \\
 185330 & 7467 &              & B5II-III   & 16167 & 3.77 &   4 & 113.4 &  7.91 & $-0.89$ &  7.02 & 360 \\
 027396 & 1350 & 53 Per       & B4IV       & 16028 & 3.91 &  15 &   5.0 &  5.27 & $-0.40$ &  4.87 & 410 \\
 034798 & 1753 &              & B3V        & 15943 & 4.27 &  37 &   7.3 &  5.53 & $-0.27$ &  5.26 & 310 \\
 184171 & 7426 &  8 Cyg       & B3IV       & 15858 & 3.54 &  27 &   7.2 &  5.39 & $-0.34$ &  5.05 & 420 \\
 198820 & 7996 &              & B3III      & 15852 & 3.86 &  32 &   8.1 &  5.49 & $-0.30$ &  5.19 & 300 \\
 030122 & 1512 &              & B5III      & 15765 & 3.72 &  15 &  87.8 &  7.51 & $-0.64$ &  6.87 & 310 \\
 020756 & 1005 &$\tau^{1}$ Ari& B5IV       & 15705 & 4.43 &  18 &   6.3 &  5.51 & $-0.29$ &  5.21 & 460 \\
 037971 & 1962 &              & B3IVp      & 15532 & 3.63 &   9 &  11.3 &  5.62 & $-0.31$ &  5.31 & 470 \\
 026739 & 1312 &              & B5IV       & 15490 & 3.92 &  31 &   7.5 &  5.47 & $-0.36$ &  5.11 & 360 \\
 209008 & 8385 & 18 Peg       & B3III      & 15353 & 3.50 &  20 &   9.9 &  5.53 & $-0.36$ &  5.17 & 340 \\
 028375 & 1415 &              & B3V        & 15278 & 4.30 &  19 &   7.4 &  5.57 & $-0.31$ &  5.27 & 430 \\
 011415 & 0542 &$\epsilon$ Cas& B3III      & 15174 & 3.54 &  42 &   9.3 &  5.50 & $-0.39$ &  5.11 & 580 \\
 147394 & 6092 & $\tau$ Her   & B5IV       & 14898 & 4.01 &  30 &   7.9 &  5.55 & $-0.37$ &  5.17 & 480 \\
 019268 & 0930 &              & B5V        & 14866 & 4.24 &  17 &   8.2 &  5.64 & $-0.31$ &  5.33 & 410 \\
 189944 & 7656 &              & B4V        & 14793 & 4.01 &  35 &   9.0 &  5.62 & $-0.35$ &  5.27 & 320 \\
 181558 & 7339 &              & B5V        & 14721 & 4.15 &  14 &  10.5 &  5.75 & $-0.29$ &  5.46 & 300 \\
 224990 & 9091 &  $\zeta$ Scl & B4III      & 14569 & 3.99 &  35 &   7.4 &  5.53 & $-0.40$ &  5.14 & 350 \\
 175156 & 7119 &              & B5II       & 14561 & 2.79 &  12 &  15.6 &  5.66 & $-0.38$ &  5.28 & 460 \\
 199578 & 8022 &              & B5V        & 14480 & 4.02 &  27 &   6.6 &  5.49 & $-0.41$ &  5.08 & 230 \\
 209419 & 8403 &              & B5III      & 14404 & 3.82 &  16 &   9.5 &  5.62 & $-0.39$ &  5.23 & 400 \\
 202753 & 8141 & 15 Aqr       & B5V        & 14318 & 3.84 &  40 &   9.5 &  5.64 & $-0.39$ &  5.25 & 360 \\
 023300 & 1141 &              & B6V        & 14207 & 3.84 &  19 &   8.6 &  5.59 & $-0.40$ &  5.19 & 540 \\
 182255 & 7358 &  3 Vul       & B6III      & 14190 & 4.29 &  28 &   4.2 &  5.40 & $-0.41$ &  4.99 & 550 \\
 041692 & 2154 &              & B5IV       & 14157 & 3.19 &  28 &  11.2 &  5.54 & $-0.46$ &  5.09 & 490 \\
 049606 & 2519 & 33 Gem       & B7III      & 14121 & 3.82 &  19 & 103.6 &  7.89 & $-0.67$ &  7.21 & 360 \\
 212986 & 8554 &              & B5III      & 14121 & 4.27 &  20 &   6.4 &  5.61 & $-0.34$ &  5.27 & 340 \\
 016219 & 0760 &              & B5V        & 14113 & 4.06 &  23 &   9.4 &  5.73 & $-0.33$ &  5.40 & 350 \\
 188892 & 7613 & 22 Cyg       & B5IV       & 14008 & 3.38 &  30 &  10.6 &  5.58 & $-0.46$ &  5.12 & 390 \\
 206540 & 8292 &              & B5IV       & 13981 & 4.01 &  13 &   7.1 &  5.58 & $-0.39$ &  5.19 & 280 \\
 191243 & 7699 &              & B5Ib       & 13923 & 2.50 &  28 &  13.7 &  5.54 & $-0.49$ &  5.05 & 410 \\
 210424 & 8452 & 38 Aqr       & B7III      & 13740 & 3.99 &  12 &   7.8 &  5.66 & $-0.38$ &  5.28 & 440 \\
 201888 & 8109 &              & B7III      & 13689 & 4.01 &   5 &   7.3 &  5.64 & $-0.38$ &  5.25 & 350 \\
 011857 & 0561 &              & B5III      & 13600 & 3.88 &  20 &   8.5 &  5.68 & $-0.40$ &  5.28 & 430 \\
 053244 & 2657 & $\gamma$ CMa & B8II       & 13467 & 3.42 &  36 &  78.4 &  7.43 & $-0.49$ &  6.94 & 510 \\
 155763 & 6396 & $\zeta$ Dra  & B6III      & 13397 & 4.24 &  41 &   9.5 &  5.92 & $-0.27$ &  5.64 & 550 \\
 173117 & 7035 &              & B5:V       & 13267 & 3.63 &  22 &   9.0 &  5.66 & $-0.47$ &  5.19 & 400 \\
 017081 & 0811 & $\pi$  Cet   & B7V        & 13063 & 3.72 &  20 &   9.9 &  5.78 & $-0.42$ &  5.37 & 510 \\
 023408 & 1149 & 20 Tau       & B8III      & 12917 & 3.36 &  30 &  69.0 &  7.33 & $-0.43$ &  6.91 & 610 \\
 196426 & 7878 &              & B8IIIp     & 12899 & 3.89 &   6 &   5.5 &  5.57 & $-0.45$ &  5.13 & 320 \\
 179761 & 7287 & 21 Aql       & B8II-III   & 12895 & 3.46 &  16 &   9.4 &  5.68 & $-0.51$ &  5.16 & 490 \\
 011529 & 0548 & $\omega$ Cas & B8III      & 12858 & 3.43 &  30 &  10.2 &  5.72 & $-0.51$ &  5.21 & 460 \\
 178065 & 7245 &              & B9III      & 12243 & 3.49 &   4 &  44.1 &  7.01 & $-0.29$ &  6.71 & 190 \\
 038899 & 2010 &134 Tau       & B9IV       & 10774 & 4.02 &  26 &   1.4 &  5.51 & $-0.33$ &  5.18 & 460 \\
 043247 & 2229 & 73 Ori       & B9II-III   & 10301 & 2.39 &  33 &   4.5 &  5.42 & $-0.67$ &  4.75 & 400 \\
 209459 & 8404 & 21 Peg       & B9.5V      & 10204 & 3.53 &   3 &   2.5 &  5.74 & $-0.33$ &  5.41 & 360 \\
 181470 & 7338 &              & A0III      & 10085 & 3.92 &   2 &   3.6 &  6.14 & $-0.19$ &  5.95 & 200 \\
\hline
\end{tabular}
\end{center}
\end{table}

%Table~1
\setcounter{table}{0}
\begin{table}[h]
\caption{(Continued.)}
%\small
\scriptsize
\begin{center}
\begin{tabular}{ccccccrrcccc}\hline\hline
\hline
HD\#  & HR\#  & star name  &  Sp.Type  & $^{*}T_{\rm eff}$  & $^{*}\log g$ & $^{\dagger}v_{\rm e}\sin i$ & $W_{6043}$ 
& $A^{\rm L}$   & $\Delta$  & $A^{\rm N}$ & $^{\#}$S/N \\
(1) & (2) & (3) & (4) & (5) & (6) & (7) & (8) & (9) & (10) & (11) & (12) \\
\hline
\multicolumn{12}{c}{[2012 May observations (late B-type stars)]}\\
 098664 & 4386 & $\sigma$ Leo & B9.5V s    & 10194 & 3.75 &  62 &  11.8 &  6.70 & $-0.14$ &  6.56 & 390 \\
 130557 & 5522 &              & B9VSi:Cr:  & 10142 & 3.85 &  55 &  10.7 &  6.69 & $-0.12$ &  6.57 & 260 \\
 079158 & 3652 & 36 Lyn       & B8IIIpMn   & 13535 & 3.72 &  46 & $\cdots$ & $\cdots$ & $\cdots$ & $\cdots$ & 230 \\
 106625 & 4662 & $\gamma$ Crv & B8IIIpHgMn & 11902 & 3.36 &  37 &  28.2 &  6.61 & $-0.35$ &  6.26 & 740 \\
 150100 & 6184 & 16 Dra       & B9.5Vn     & 10542 & 3.84 &  36 &   3.2 &  5.89 & $-0.25$ &  5.64 & 320 \\
 197392 & 7926 &              & B8II-III   & 13166 & 3.46 &  30 &  11.3 &  5.74 & $-0.49$ &  5.25 & 280 \\
 198667 & 7985 &  5 Aqr       & B9III      & 11125 & 3.42 &  26 &   5.0 &  5.73 & $-0.40$ &  5.33 & 220 \\
 202671 & 8137 & 30 Cap       & B8III      & 13566 & 3.36 &  25 &  85.7 &  7.52 & $-0.55$ &  6.97 & 220 \\
 193432 & 7773 &  $\nu$ Cap   & B9.5V      & 10180 & 3.91 &  24 &   3.2 &  6.05 & $-0.21$ &  5.84 & 430 \\
 161701 & 6620 &              & B9pHgMn    & 12692 & 4.04 &  20 &  48.2 &  7.24 & $-0.18$ &  7.06 & 310 \\
 077350 & 3595 & $\nu$ Cnc    & A0pSi      & 10141 & 3.68 &  20 &   3.5 &  6.01 & $-0.25$ &  5.75 & 310 \\
 129174 & 5475 & $\pi^{1}$ Boo& B9pMnHgSi  & 12929 & 4.02 &  16 &  26.3 &  6.60 & $-0.23$ &  6.36 & 370 \\
 201433 & 8094 &              & B9VpSi     & 12193 & 4.24 &  15 &   3.1 &  5.59 & $-0.35$ &  5.24 & 190 \\
 144206 & 5982 &$\upsilon$ Her& B9III      & 11925 & 3.79 &  12 &  14.5 &  6.30 & $-0.26$ &  6.03 & 450 \\
 145389 & 6023 & $\phi$ Her   & B9p:Mn:    & 11714 & 4.02 &  11 &   2.3 &  5.47 & $-0.41$ &  5.06 & 490 \\
 190229 & 7664 &              & B9pHgMn    & 13102 & 3.46 &  10 &  73.6 &  7.43 & $-0.43$ &  6.99 & 280 \\
 149121 & 6158 & 28 Her       & B9.5III    & 10748 & 3.89 &  10 &   8.6 &  6.38 & $-0.17$ &  6.21 & 420 \\
 078316 & 3623 & $\kappa$ Cnc & B8IIIpMn   & 13513 & 3.85 &   8 & 100.7 &  7.94 & $-0.64$ &  7.30 & 290 \\
 089822 & 4072 &              & A0pSiSr:Hg:& 10307 & 3.89 &   5 &   6.4 &  6.36 & $-0.16$ &  6.20 & 400 \\
 143807 & 5971 & $\iota$ CrB  & A0p:Hg:    & 10828 & 4.06 &   4 &   8.9 &  6.45 & $-0.14$ &  6.31 & 470 \\
 193452 & 7775 &              & A0III      & 10543 & 4.15 &   3 &   3.7 &  6.11 & $-0.17$ &  5.93 & 170 \\
\hline
\end{tabular}
\end{center}
(1) Henry Draper Catalogue number. (2) Bright Star Catalogue number (Hoffleit \& Jaschek 1991).
(3) Star name in the constellation. (4) Spectral type (taken from Hoffleit \& Jaschek 1991).
(5) Effective temperature (in K). (6) Logarithmic surface gravity (in cm~s$^{-2}$/dex).
(7) Projected rotational velocity (in km~s$^{-1}$). (8) Equivalent width of P~{\sc ii} 6043 line
(in m\AA). (9) P abundance in LTE (in dex). (10) Non-LTE correction (in dex). 
(11) P abundance in non-LTE (in dex). (12) Signal-to-noise ratio. \\
The first part present the data of 64 early-to-late B stars (observed in 2006 October),
followed by the second part for 21 late B-type stars (observed in 2012 May).
The former is arranged in the descending order of $T_{\rm eff}$ while the latter 
is in the descending order of $v_{\rm e}\sin i$, in order to keep consistency with the 
corresponding original papers (Paper~I and Paper~II). \\
$^{*}$Determined from color indices of Str\"{o}mgren's $uvby\beta$ photometry  (cf. Sect.~3).\\
$^{\dagger}$Determined from the fitting analysis around $\sim$~6150--6160~\AA\ region 
(cf. Paper~I and Paper~II).\\
$^{\#}$Mean of the results measured from five segments in the neighborhood of the P~{\sc ii} 6043 line. 
\end{table}

%Fig. 2
\begin{figure}[h]
\begin{minipage}{150mm}
\begin{center}
\includegraphics[width=6.0cm]{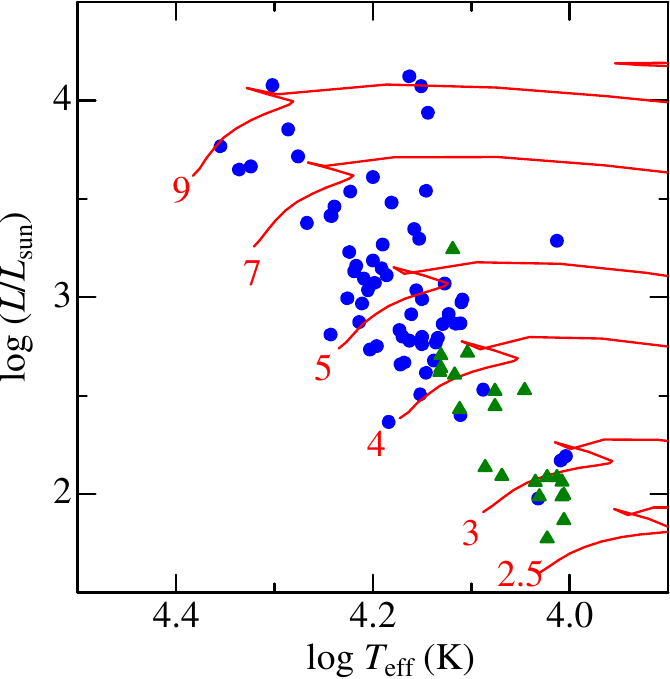}
\end{center}
\FigCap{ 
Plots of 85 program stars on the theoretical HR diagram 
($\log (L/L_{\odot})$ vs. $\log T_{\rm eff}$), where blue filled circles
and green filled triangles correspond to 64 stars of 2006 October 
observations and 21 stars of 2012 May observations, respectively. 
Theoretical evolutionary tracks of solar metallicity stars 
for six different initial masses (2.5, 3, 4, 5, 7, and 9 $M_{\odot}$),  
which were computed by Lejeune \& Schaerer (2001), are also depicted by red lines 
for comparison. See Sect.~3 for the adopted $T_{\rm eff}$, and the caption 
of Fig.~1 in Paper~I (or Paper II) regarding how $L$ was evaluated. 
}
\end{minipage}
\end{figure}

\section{Atmospheric Parameters}

Regarding the effective temperature ($T_{\rm eff}$) and the surface gravity ($\log g$)
of each star, the same values as adopted in Papers I and II are used unchanged (cf. Table~1),
which were determined from colors ($b-y$, $c_{1}$, $m_{1}$, and $\beta$) 
of Str\"{o}mgren's $uvby\beta$ photometric system. 

As to the microturbulence ($\xi$), the values assumed in Paper I 
[$3 (\pm 2)$~km~s$^{-1}$ for early-to-late B-type stars] and Paper~II 
[$1 (\pm 1)$~km~s$^{-1}$ for late B-type stars] were not consistent.
In this paper, $T_{\rm eff}$-dependent values are assigned as 
$\xi = 1 (\pm 1)$~km~s$^{-1}$ (10000~K~$< T_{\rm eff} < 16500$~K) and
$\xi = 2 (\pm 1)$~km~s$^{-1}$ (16500~K~$< T_{\rm eff} < 23000$~K).

This is due to the fact that $\xi$ plays a significant role only for the case of
P-rich chemically peculiar stars (HgMn stars) found only among late B-type stars, 
while the P-line strengths of normal stars (existing over the entire $T_{\rm eff}$ 
range) are so weak that their P abundances are practically $\xi$-independent. 
Therefore, the same value as in Paper~II ($\xi = 1$~km~s$^{-1}$; specific 
to late B-type stars) is adopted for $T_{\rm eff} < 16500$~K. while a tentative 
value of 2~km~s$^{-1}$ is roughly assumed at $T_{\rm eff} > 16500$~K 
(all are normal stars with weak lines of phosphorus). 

The model atmospheres adopted for each of the targets are the same as in Papers I 
and II, which are the solar-metallicity models constructed by two-dimensionally 
interpolating Kurucz's (1993) ATLAS9 model grid in terms of $T_{\rm eff}$ and $\log g$.

\section{Spectrum Fitting Analysis and Evaluation of Equivalent Widths}

As already mentioned in Sect.~1, it is important to employ a spectral line 
of as large transition probability ($\log gf$) as possible for successful 
P abundance determinations of B-type stars, because lines tend to be 
considerably weak and hard to detect for near-normal P abundances.
In this respect, the suitable candidate lines in the optical region 
are the P~{\sc ii} lines of 4s\,$^{3}$P$^{\circ}$\,--\,4p\,$^{3}$D transition 
(lower excitation potentials of $\chi_{\rm low} \sim $\,10.8\,eV, multiplet~5). 
They are at 6024.13, 6034.34, 6043.08, 6087.84, and 6165.60~\AA\  
and have $\log gf$ values of +0.20, $-0.15$, $+0.44$, $-0.38$, and $-0.41$,
respectively, according to the VALD database (Ryabchikova et al. 2015). 
Therefore, we invoke the P~{\sc ii} line at 6043.08~\AA, which is the strongest 
one among these and also free from any appreciable blending. Besides, 
this line has another advantage that it lies almost in the middle part of 
the relevant Echelle order (covering 5990--6100~\AA) where the S/N ratio 
is comparatively higher. 

The procedures of analysis (spectrum fitting, equivalent width derivation, 
estimating abundance errors due to parameter uncertainties) are essentially the same 
as adopted in Papers I and II (cf. Sect.~4 therein). Note that all the calculations 
in this section are done with the assumption of LTE at this stage. 

\subsection{Synthetic Spectrum Fitting}

First, a spectrum-fitting technique was applied to the 6040--6050~\AA\ region
(comprising the P~{\sc ii} 6043 line), by which the best-fit between theoretical and 
observed spectra is accomplished. Here, the parameters varied are the abundances of 
P, O, and Ne (+ Mn if necessary), rotational broadening velocity ($v_{\rm e}\sin i$), 
and radial velocity ($V_{\rm rad}$). The data of all atomic lines included in this
wavelength region were taken from the VALD database (Ryabchikova et al. 2015).  
Specifically, the data for the relevant P~{\sc ii} line at 6043.084~\AA\ are 
$\chi_{\rm low} = 10.802$~eV (lower excitation potential), $\log gf = +0.442$
(logarithmic $gf$ value), Gammar = 9.22 (radiation damping parameter), 
Gammas = $-5.76$ (Stark effect damping parameter), and Gammaw = $-7.73$ (van der Waals 
effect damping parameter).\footnote{
Gammar is the radiation damping width (s$^{-1}$), $\log\gamma_{\rm rad}$.
Gammas is the Stark damping width (s$^{-1}$) per electron density (cm$^{-3}$) 
at $10^{4}$ K, $\log(\gamma_{\rm e}/N_{\rm e})$. 
Gammaw is the van der Waals damping width (s$^{-1}$) per hydrogen density 
(cm$^{-3}$) at $10^{4}$ K, $\log(\gamma_{\rm w}/N_{\rm H})$.}
The phosphorus abundances could be successfully established for 83 stars 
(out of 85 program stars), except for HD~029248 and HR~3652, for which 
the P abundance was tentatively fixed at an arbitrary value in the fitting.
The agreement between the theoretical spectrum (for the solutions 
with converged parameters) with the observed spectrum for each star 
is shown in Fig.~3.

%Fig. 3
\begin{figure}[h]
\begin{minipage}{150mm}
\begin{center}
\includegraphics[width=12.0cm]{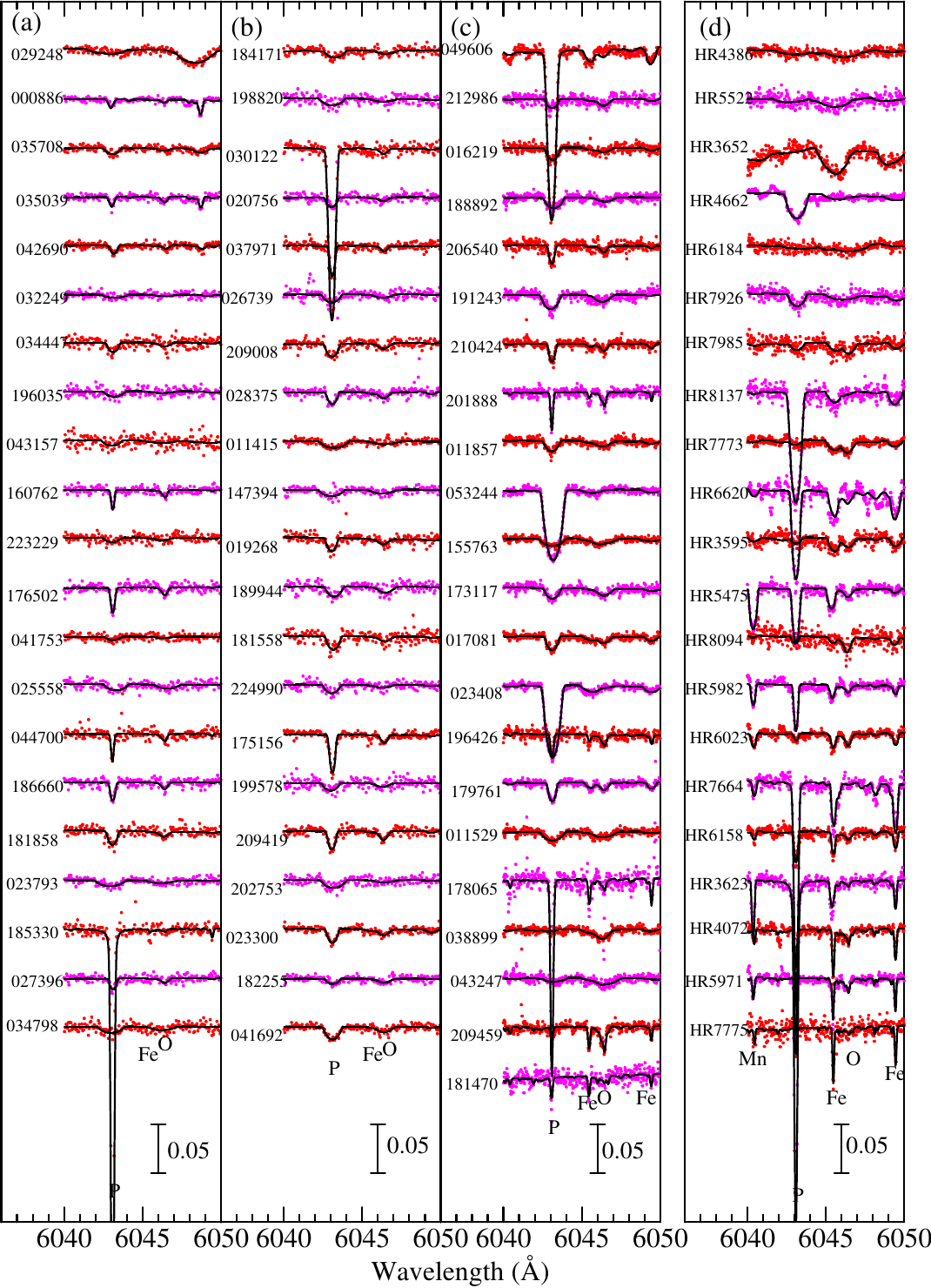}
\end{center}
\FigCap{
Synthetic spectrum fitting at the 6040--6050~\AA\ region for the determination
of P abundances. The best-fit theoretical spectra are depicted by solid lines, 
while the observed data are plotted by dots.  
The spectra are arranged in the same manner as in Table~1:  
Panels (a), (b), and (c) are for the 64 early-to-late B stars of 2006 October 
observations (indicated by the HD number) in the descending order of 
$T_{\rm eff}$ (from top to bottom; from left to right). while the rightmost 
panel (d) is for 21 late-B stars of 2012 May observations (indicated by the 
HR number) in the descending order of $v_{\rm e}\sin i$. An offset of 0.05 
(in unit of the continuum-normalized flux) is applied to each spectrum 
relative to the adjacent one. 
}
\end{minipage}
\end{figure}

\subsection{Equivalent Widths and Their Errors}

Next, the equivalent width of the P~{\sc ii} 6043 line $W_{6043}$ was inversely 
evaluated from the abundance solution resulting from the fitting analysis. 
The $W_{6043}$ values derived in this manner are given in Table~1. While 
this line is generally weak for ordinary B stars (1~m\AA $\lesssim W_{6043} \lesssim 10$~m\AA), 
it can be stronger (up to $W_{6043} \sim 100$~m\AA) for P-rich peculiar stars (cf. Fig.~4a). 

The error involved with $W_{6043}$ was estimated as
\begin{equation}
\delta W \sim \epsilon W_{6043}(1 - R_{0})/R_{0}
\end{equation}
according to Takeda (2023; cf. Sect.~6.2 therein) 
where $\epsilon (\equiv ({\rm S/N})^{-1})$ is the random fluctuation of the 
continuum level and $R_{0} (\equiv 1 - F_{0}/F_{\rm c})$ is the line depth 
at the line center. 
By inserting the S/N ratios (measured around the P~{\sc ii} 6043 line; see column~12 
in Table~1) into Eq.~(1), typical values of $\delta W$ were found to be 
$\sim$~1--3 m\AA\ in most cases\footnote{
Errors of $W_{6043}$ evaluated by Cayrel's (1988) formula (depending on S/N, pixel 
size, and line widths) were found to be smaller (typically by several times) than 
$\delta W$ defined by Eq.~(1), and thus not taken into account.} (or up to 
$\sim 10$~m\AA\ in the exceptional case of low S/N and very shallow $R_{0}$), 
as depicted by error bars attached to the symbols in Fig.~4a.
The impact of $\delta W$ on the P abundance (denoted as $\delta_{W}$) can be significant 
(e.g., a few tenths dex or even more) for the very weak line case where $W_{6043}$ 
and $\delta W$ are on the similar size.

%Fig. 4
\begin{figure}[h]
\begin{minipage}{150mm}
\begin{center}
\includegraphics[width=9.0cm]{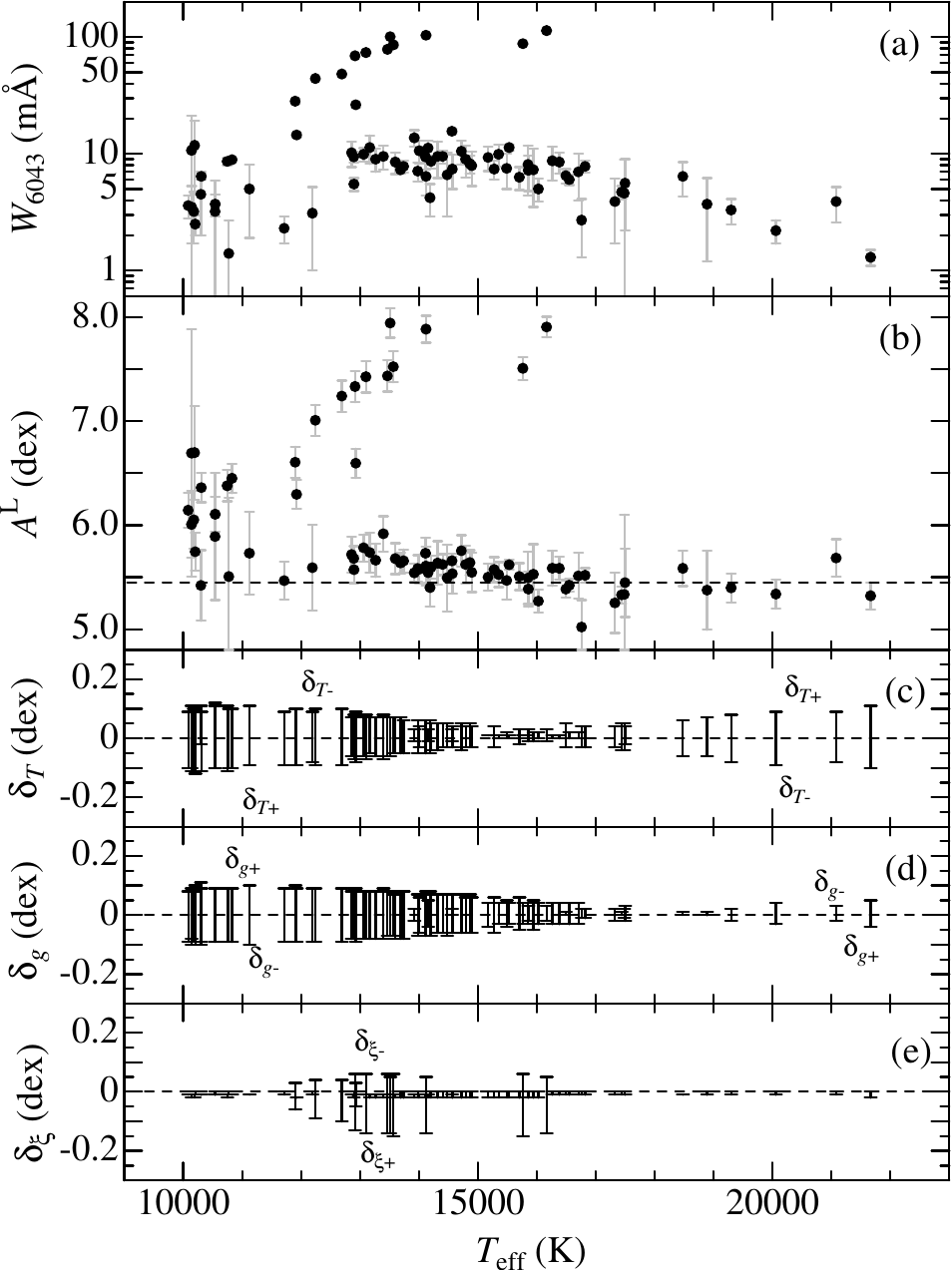}
\end{center}
\FigCap{
The equivalent widths of the P~{\sc ii}~6043 line, resulting LTE abundances,
and sensitivities due to perturbations of atmospheric parameters, 
are plotted against $T_{\rm eff}$.
(a) $W_{6043}$ (equivalent width), where the indicated error bars ($\delta W$) 
are their uncertainties (see Sect.~4.2 for more details).
(b) $A^{\rm L}$ (LTE phosphorus abundance).
Here, the attached error bars are the root-sum-squares of $\delta_{W}$
(abundance ambiguities corresponding to $\delta W$), 
$\delta_{T}$, $\delta_{g}$, and $\delta_{\xi}$.
(c) $\delta_{T+}$ and $\delta_{T-}$ (abundance variations 
in response to $T_{\rm eff}$ changes of $+3 \%$ and $-3 \%$). 
(d) $\delta_{g+}$ and $\delta_{g-}$ (abundance variations 
in response to $\log g$ changes of $+0.2$~dex and $-0.2$~dex). 
(e) $\delta_{\xi +}$ and $\delta_{\xi -}$ (abundance variations 
in response to perturbing the standard $\xi$ value by $\pm 1$~km~s$^{-1}$). 
Note that the signs of $\delta_{T}$ and $\delta_{g}$ are reversed on 
both sides of $T_{\rm eff}$ around $\sim$~16000--18000~K (mid B-type).
}
\end{minipage}
\end{figure}

\subsection{Impact of Parameter Uncertainties}

How the ambiguities in atmospheric parameters ($T_{\rm eff}$, $\log g$, 
and $\xi$) affect the P abundances was estimated by repeating the analysis on 
the $W_{6043}$ values while perturbing these standard parameters interchangeably 
by $\pm 3\%$, $\pm 0.2$~dex, and $\pm 1$~km~s$^{-1}$, which are the typical 
uncertainties for $T_{\rm eff}$ and $\log g$ (cf. Sect.~3 in Paper~I) and 
for $\xi$ (cf. Sect.~3 of this paper).

The resulting $A^{\rm L}$ (P abundances in LTE), 
$\delta_{T\pm}$ (abundance changes by perturbations of $T_{\rm eff}$), 
$\delta_{g\pm}$ (abundance changes by perturbations of $\log g$), and
$\delta_{\xi\pm}$ (abundance changes by perturbations of $\xi$)
are plotted against $T_{\rm eff}$ in Figs.~4b, 4c, 4d, and 4e,
respectively. 

As seen from Figs.~4c and 4d, both $|\delta_{T}|$ and $|\delta_{g}|$ 
are $\lesssim 0.1$~dex and thus not very significant.
According to Fig.~4e, $|\delta_{\xi}|$ is negligibly small for normal stars,
while $|\delta_{\xi}|$ amounts up to $\sim 0.1$~dex for P-rich late B-type 
peculiar stars. The error bars attached to the symbols in Fig.~4b
are the root-sum-squares of $\delta_{W}$, $\delta_{T}$, $\delta_{g}$,
and $\delta_{\xi}$.

\section{Statistical Equilibrium Calculations for P~{\sc ii}}

\subsection{Atomic Model}

The non-LTE calculations for P~{\sc ii} were carried out based on the P~{\sc ii} model 
atom comprising 83 terms (up to 3s$^{2}$\,3p\,9g at 154582 cm$^{-1}$) and 1206 radiative 
transitions, which was constructed by consulting the updated atomic line data 
(filename: ``gfall21oct16.dat'')\footnote{http://kurucz.harvard.edu/linelists/gfnew/}
 compiled by Dr. R. L. Kurucz. Though the contribution of P~{\sc i} was neglected,
P~{\sc iii} was taken into account in the number conservation of total P atoms. 

Regarding the photoionization cross section, the data calculated by Nahar et al. 
(2017)\footnote{The cross-section profiles (as functions of ionizing photon energy)
were roughly digitized from their Fig.~1, where attention was paid to reproduce 
only the global trend, because many sharp resonance peaks included in the 
original data are very difficult to read out.}
were used for the lowest 4 terms ($^{3}$P, $^{1}$D, $^{1}$S, and $^{5}$S$^{\circ}$), 
while the hydrogenic approximation was assumed for the remaining terms.  
Otherwise (such as the treatment of collisional rates), the recipe described in Sect.~3.1.3 
of Takeda (1991) was followed (inelastic collisions due to neutral hydrogen 
atoms were formally included as described therein, though insignificant 
in the atmosphere of early-type stars considered here).   

\subsection{Grid of Models}

The calculations were done on a grid of 36 ($= 9 \times 4$) 
solar-metallicity model atmospheres 
resulting from combinations of nine $T_{\rm eff}$ values 
(9000, 10000, 12000, 14000, 16000, 18000, 20000, 22000, and 24000~K) 
and four  $\log g$ values (3.0, 3.5, 4.0, and 4.5).
while assuming $\xi$ = 2~km~s$^{-1}$.
Regarding the input P abundance, three values of $A$(P) = 4.45,
5.45, and 6.45 (corresponding to [P/H] = $-1$, 0, and +1) were assumed, 
resulting in three kinds of non-LTE grids.  
The depth-dependent non-LTE departure coefficients to be used for each star were 
then evaluated by interpolating the grid (for each [P/H]) 
in terms of $T_{\rm eff}$ and $\log g$.

\section{Non-LTE Effect on P Abundance Determinations}

\subsection{Characteristic Trends}

%Fig. 5
\begin{figure}[h]
\begin{minipage}{150mm}
\begin{center}
\includegraphics[width=9.0cm]{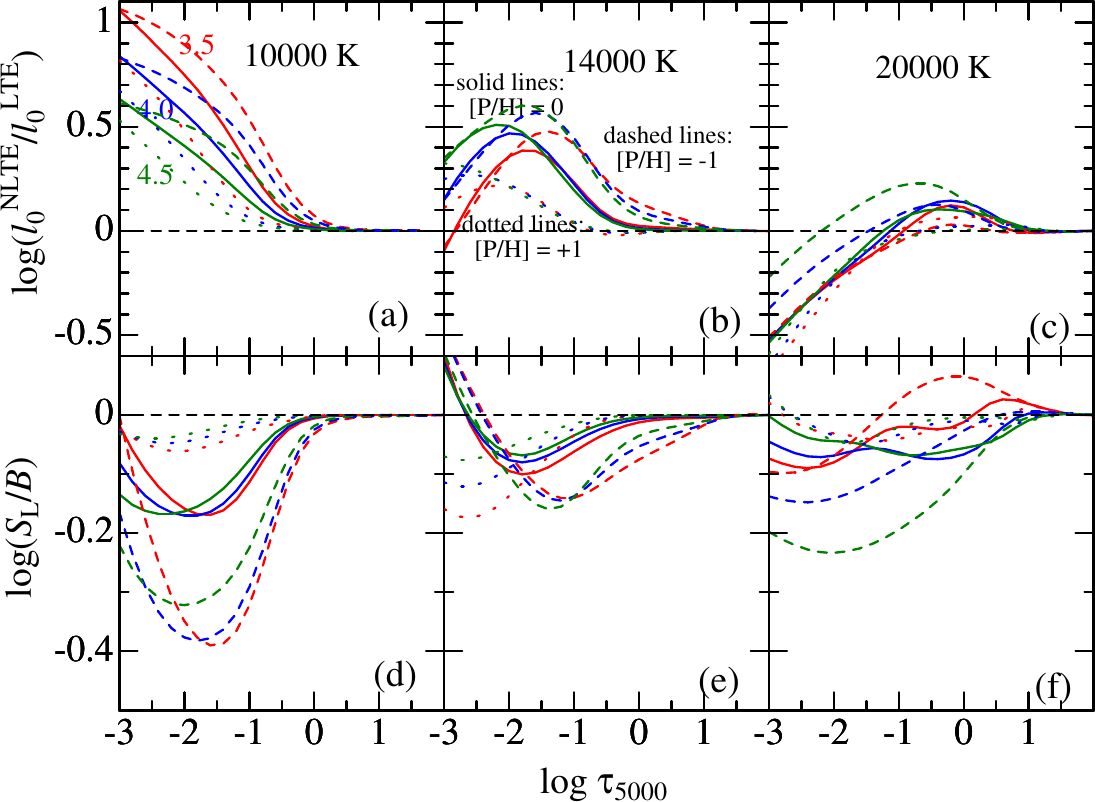}
\end{center}
\FigCap{
The non-LTE-to-LTE line-center opacity ratio (upper panels (a)--(c)) and 
the ratio of the line source function ($S_{\rm L}$) 
to the local Planck function ($B$) (lower panels (d)--(f))  
for the P~{\sc ii} 4s\,$^{3}$P$^{\circ}$\,--\,4p\,$^{3}$D transition 
(corresponding to P~{\sc ii} 6043.08 line) of multiplet~5, 
plotted against the continuum optical depth at 5000~\AA. 
Shown here are the calculations done with $\xi = 2$~km~s$^{-1}$ 
on the solar-metallicity models of 
$T_{\rm eff} =$ 10000~K (left panels (a), (d)), 14000~K (middle panels (b), (e)), 
and 20000~K (right panels (c), (f)). At each panel, the results for three 
different P abundances ([P/H] = $-1$, 0, and +1) are discriminated by 
line-types (dashed, solid, and dotted lines, respectively), while those 
for three $\log g$ values of 3.5, 4.0, and 4.5 are depicted by different 
colors (red, blue, and green, respectively). 
}
\end{minipage}
\end{figure}

%Fig. 6
\begin{figure}[h]
\begin{minipage}{150mm}
\begin{center}
\includegraphics[width=9.0cm]{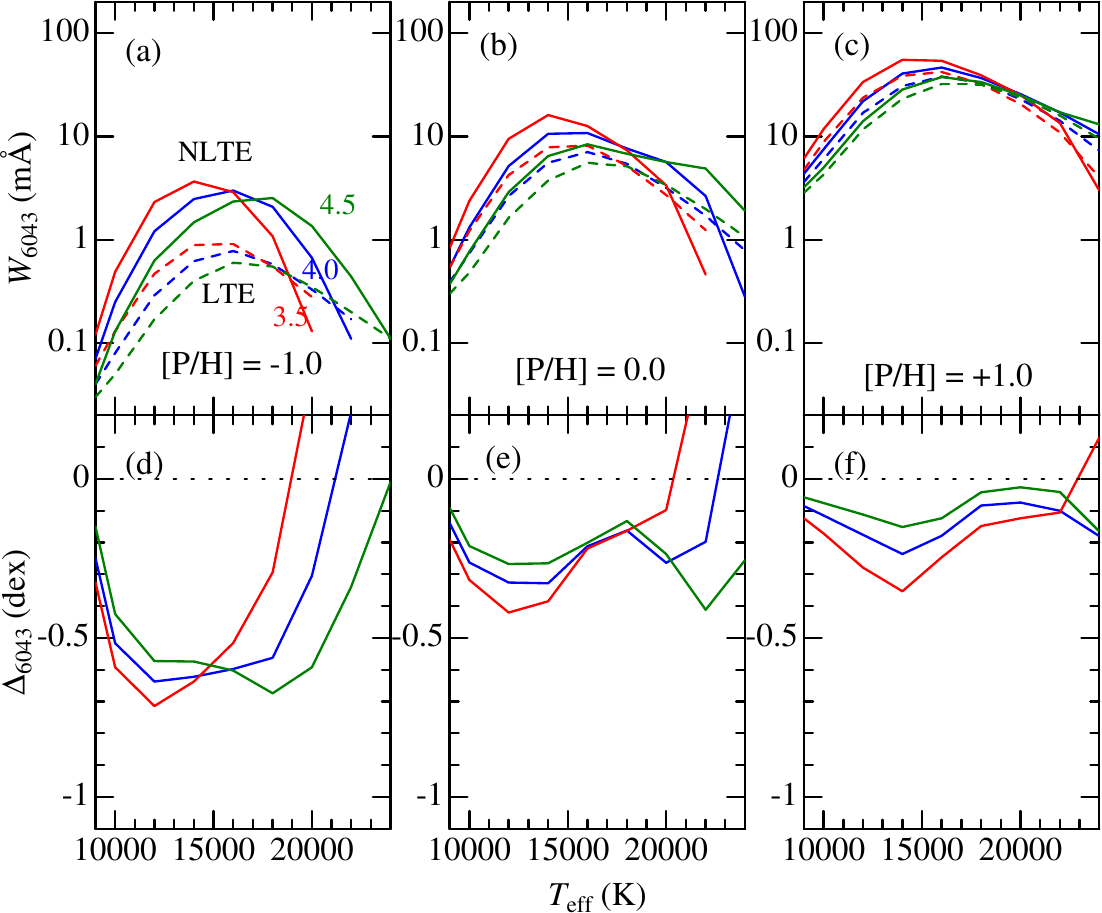}
\end{center}
\FigCap{
The non-LTE and LTE equivalent widths ($W^{\rm N}$ and $W^{\rm L}$) 
for the P~{\sc ii} 6043 line and the corresponding non-LTE corrections 
($\Delta \equiv A^{\rm N} - A^{\rm L}$, where $A^{\rm L}$ and $A^{\rm N}$ are 
the abundances derived from $W^{\rm N}$ with LTE and non-LTE), which were 
computed on the non-LTE grid of models described in 
Sect.~5.2, are plotted against $T_{\rm eff}$.
The upper panels ((a)--(c)) are for $W^{\rm N}$ (solid lines) and $W^{\rm L}$ 
(dashed lines), while the lower panels ((d)--(f)) are for $\Delta$.
The left ((a), (d)), middle ((b), (e)), and right ((c), (f)) panels 
correspond to [P/H] = $-1$, 0, and +1, respectively.
In each panel, the results for different $\log g$ (3.5, 4.0, and 4.5)
are depicted in different colors (red, blue, and green, respectively), 
}
\end{minipage}
\end{figure}

Fig.~5 displays the $l_{0}^{\rm NLTE}(\tau)/l_{0}^{\rm LTE}(\tau)$ (the 
non-LTE-to-LTE line-center opacity ratio; almost equal to $\simeq b_{\rm l}$) and 
$S_{\rm L}(\tau)/B(\tau)$ (the ratio of the line source function to the Planck 
function; nearly equal to $\simeq b_{\rm u}/b_{\rm l}$) for the transition 
relevant to the P~{\sc ii} 6043 line ($b_{\rm l}$ and $b_{\rm u}$ are the non-LTE 
departure coefficients for the lower and upper levels) as functions of optical depth 
at 5000\AA\ for selected representative cases. 
Likewise, Fig.~6 illustrates how the theoretical equivalent widths calculated in LTE 
($W^{\rm L}$) as well as in non-LTE ($W^{\rm L}$) and the corresponding non-LTE 
corrections ($\Delta \equiv A^{\rm N} - A^{\rm L}$, where $A^{\rm L}$ and $A^{\rm N}$ are 
the abundances derived from $W^{\rm N}$ with LTE and non-LTE) depend upon $T_{\rm eff}$.
The following characteristic trends are read from these figures.
\begin{itemize}
\item
As seen from Fig.~6, the inequality $W^{\rm N} > W^{\rm L}$ (and $\Delta < 0$) holds 
in most cases (except for the high $T_{\rm eff}$ end at $\gtrsim 20000$~K where the trend
is inverse), which means that the non-LTE effect tends to strengthen the 
P~{\sc ii} 6043 line. 
\item
For a given [P/H], $|\Delta|$ almost correlates with $W$ as well as $T_{\rm eff}$ 
in the sense that $|\Delta|$ takes the largest value around $T_{\rm eff} \sim$~15000~K 
where $W$ reaches a maximum, as recognized from each panel of Fig.~6. 
This is understandable because an increase of $W$ makes the line formation zone shallower 
where the departure from LTE is larger. 
\item
As to $g$-dependence of the non-LTE effect, $|\Delta|$ tends to be larger for lower 
$\log g$ (i.e., lower density atmosphere) as seen at $T_{\rm eff} \lesssim 15000$~K, 
though the situation is not so simple at the higher $T_{\rm eff}$ regime
where $\Delta$ gradually shifts in the direction of changing its sign. 
\item
A more important factor significantly affecting the degree of departure from LTE is 
the P abundance ([P/H]) assumed in the calculations. That is, the non-LTE effect 
($|\Delta|$) tends to be progressively less significant with an increase in [P/H] 
(Fig.~6d $\rightarrow$ Fig.~6e $\rightarrow$ Fig.~6f) if compared at the same 
$T_{\rm eff}$ and $\log g$, despite that the line strength ($W$) is enhanced 
with increasing [P/H] (Fig.~6a $\rightarrow$ Fig.~6b $\rightarrow$ Fig.~6c).
\item
This [P/H]-dependence may be understood by considering the mechanism controlling 
the level populations for the relevant P~{\sc ii} 6043 line.
The lower level (4s\,$^{3}$P$^{\circ}$; $\chi_{\rm low} \simeq 10.8$~eV) 
of this transition is radiatively connected to the ground level (3p$^{2}$\,$^{3}$P; 
$\chi_{\rm low} \simeq 0$~eV) by lines of large transition probabilities at 
$\sim$~1150--1160~\AA. Then, this strong UV transition is almost in the condition 
of radiative detailed balance (if it is sufficiently optically thick), which 
makes the 4s\,$^{3}$P$^{\circ}$ level as if ``meta-stable''.
In this case, if the subordinate lines originating from 4s\,$^{3}$P$^{\circ}$ 
become optically thin, this level would be overpopulated ($b > 1$) by cascading 
from the upper levels. Here, the P abundance would play a significant role 
in the sense that lower [P/H] (thinner optical thickness of subordinate lines) 
leads to more enhanced cascades and larger overpopulation. Actually, the extent 
of overpopulation systematically increases with a decrease in [P/H] if compared
at the same depth (dotted lines $\rightarrow$ solid lines $\rightarrow$ 
dashed lines in Fig.~5a--5c).
\item
Note, however, that this argument is based on the presumption that
the 3p$^{2}$\,$^{3}$P\,--\,4s\,$^{3}$P$^{\circ}$ UV transition is so optically 
thick to be in detailed balancing. This condition is destined to break down 
when double ionization proceeds and P~{\sc ii} is replaced by P~{\sc iii} 
($T_{\rm eff} \gtrsim 16000$~K; cf. Fig.~1d), because the population of 
P~{\sc ii} ground level comes short due to ionization. In this case,
the 4s\,$^{3}$P$^{\circ}$ level can not be meta-stable any more, and 
eventually becomes underpopulated ($b < 1$). Figs.~5a--5c illustrate this 
situation of how the overpopulation progressively turns into underpopulation 
in the optically-thin layer as $T_{\rm eff}$ increases from late B to early B. 
This also explains the reason why $\Delta$ becomes positive (i.e., non-LTE
line weakening) at $T_{\rm eff} \gtrsim 20000$~K (cf. Figs.~6d--6f). 
\end{itemize} 

\subsection{Non-LTE Corrected Abundances}

Since non-LTE corrections are appreciably dependent upon [P/H], it is necessary 
to adopt a correction corresponding to an adequate [P/H] consistent with 
the final non-LTE abundance. Therefore, we proceed as follows.

First, three kinds of non-LTE abundances are derived from $W_{6043}$ 
($A_{-1}^{\rm N}$, $A_{0}^{\rm N}$, and $A_{+1}^{\rm N}$, corresponding to
[P/H] = $-1$, 0, and +1, respectively) by using three sets of departure 
coefficients prepared for each star (cf. Sect.~5.2).
These three non-LTE abundances are sufficient to express $A^{\rm N}$
by a second-order polynomial in terms of $x (\equiv {\rm [P/H]})$ as 
\begin{equation}
A^{\rm N} = a x^{2} + b x + c,
\end{equation} 
where $a$, $b$, and $c$ are known coefficients.
Meanwhile, according to the definition,
\begin{equation}
A^{\rm N} = x + 5.45.  
\end{equation}
Combining Eqs.~(2) and (3), we have 
\begin{equation}
a x^{2} + b x + c = x+ 5.45.
\end{equation} 
Let us denote the solution of Eq.~(4) as $x_{*}$ (which of two solutions  
should be adopted is self-evident), from which we obtain 
$A_{*}^{\rm N} (= x_{*} + 5.45)$ and 
$\Delta_{*} (= A_{*}^{\rm N} - A^{\rm L})$   
as the final non-LTE abundance and non-LTE correction.

Such derived $\Delta_{*}$ values are plotted against $T_{\rm eff}$ in Fig.~7b 
(black filled circles), where $\Delta_{-1}$ (blue), $\Delta_{0}$ (green), 
and $\Delta_{+1}$ (red) are also overplotted by open symbols for comparison.
Likewise, the final non-LTE abundances ($A_{*}^{\rm N}$) are shown 
against $T_{\rm eff}$ in Fig.~7c.
 
%Fig. 7
\begin{figure}[h]
\begin{minipage}{150mm}
\begin{center}
\includegraphics[width=9.0cm]{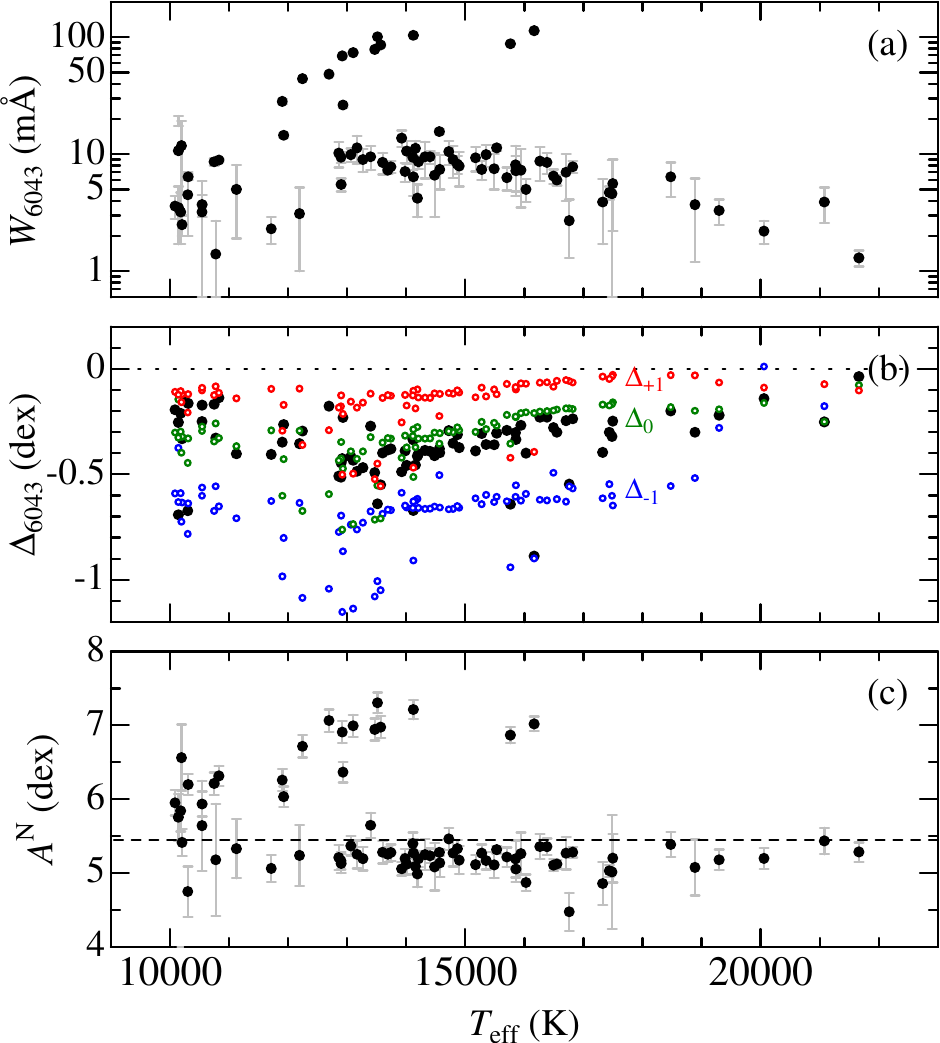}
\end{center}
\FigCap{
The equivalent widths (same as in Fig.~4a) of the P~{\sc ii} 6043 line, 
the finally adopted non-LTE corrections, and the corresponding non-LTE abundances 
for each of the program stars are plotted by black filled circles against 
$T_{\rm eff}$ in panels (a), (b), and (c), respectively..
In panel (b), the original non-LTE corrections ($\Delta_{-1}$, $\Delta_{0}$, 
and $\Delta_{+1}$; corresponding to [P/H] = $-1$, 0, and +1), 
from which the final $\Delta$ was derived, are
also depicted for comparison by open circles (colored in blue, green, and 
red, respectively). The error bars attached to the symbols in panels (a) 
and (c) are the same as in Figs.~4a and 4b.  
}
\end{minipage}
\end{figure}

\section{Phosphorus Abundances of B-type Stars}

It is apparent from Fig.~7c that the P abundances of the program stars are divided into two 
groups: (i) late B-type chemically peculiar stars which exhibit conspicuous overabundances 
systematically increasing from $A^{\rm N} \sim 6$ ($T_{\rm eff} \sim 10000$~K) to 
$A^{\rm N} \sim 7$ ($T_{\rm eff} \sim 16000$~K), and (ii) normal early-to-late B-type
stars ($10000 \lesssim T_{\rm eff} \lesssim 22000$~K) which show rather
similar P abundances irrespective of $T_{\rm eff}$. These two groups of stars are 
separately discussed below. 

\subsection{P-rich Peculiar Stars}

The former P-enhanced group mostly consists of non-magnetic late B-type 
chemically peculiar stars (HgMn stars), which are known to show considerable
overabundance of P.  Nevertheless, P-rich stars and those classified as 
HgMn peculiar stars are not strictly equivalent but some exceptions do exist
(P-strong stars classified as normal or P-weak HgMn stars).

The tendency observed in Fig.~7c (P is overabundant in HgMn stars and its anomaly
increases with $T_{\rm eff}$) is actually a reconfirmation of the trend shown 
in Fig.~4 of Ghazaryan \& Alecian (2016), though the extent of overabundance
seen in their figure ([P/H] $\sim +2$ at $T_{\rm eff} \sim$~14000~K) appears to 
be somewhat overestimated by $\sim +0.5$~dex in comparison with Fig.~7c,
which is presumably due to their neglect of non-LTE corrections.

It should be noted, however, that P abundances derived by using the conventional 
model atmosphere (1D plane-parallel model with vertically changing physical variables 
but homogeneous abundances) are of limited significance in the present case 
of HgMn stars, because the chemical composition of P is likely to be stratified in their 
atmospheres (i.e., increasing with height) due to the element segregation process 
as indicated by recent studies (see, e.g., Catanzaro et al. 2016, Ndiaye et al. 2018, 
Alecian \& Stift 2019).

\subsection{Superficially Normal B-type Stars}

We are now to discuss the photospheric P abundances of normal B-type stars, 
which should retain the composition of galactic gas from which they were formed.

As long as LTE abundances are concerned, their $A^{\rm L}$ values almost
distribute around the solar abundance ($A_{\odot} = 5.45$) but exhibit a
$T_{\rm eff}$-dependent trend at $T_{\rm eff} \lesssim 16000$~K 
(d$A^{\rm L}$/d$T_{\rm eff} \sim$~0.1~dex/1000~K) as shown in Fig.~4c.

However, since the non-LTE corrections ($\Delta$) also have a systematic gradient with 
$T_{\rm eff}$ just in the inverse sense (Fig.~7b), the non-LTE abundances ($A^{\rm N}$) 
turn out to be almost independent upon $T_{\rm eff}$ (Fig.~7c), accomplishing a reasonable
homogeneity. An inspection of Fig.~7c suggests that the demarcation line dividing
the two groups may be set at $\sim 5.7$. Then, those 61 stars satisfying the criterion 
$A^{\rm N} < 5.7$ are regarded as normal B stars, for which the mean abundance is 
calculated as $\langle A^{\rm N} \rangle = 5.20$ (standard deviation is $\sigma =  0.18$).
   
Here, we are confronted with a somewhat puzzling problem if this result is compared 
with the reference solar abundance of $A_{\odot} = 5.45$. That is, P abundances 
of young B-type stars (representing the gas composition at the time of 
some $\sim 10^{7}$--$10^{8}$~yr ago) are by $\sim$~0.2--0.3~dex ``lower''
than that of the Sun (formed $\sim 4.6 \times 10^{9}$~yr ago).
This may suggests that P abundance of galactic gas has ``decreased'' with time, 
which apparently contradicts the standard concept of chemical evolution in the Galaxy
(elements are synthesized and expelled by stars, by which gas is chemically 
enriched with the lapse of time).  

Given that P abundances were determined based only on one line (P~{\sc ii} 6043), 
whether its transition probability is credible or not may be worth checking, 
because it directly affects the result. The adopted $\log gf  = +0.442$ for 
this line (taken from VALD) was obtained by Dr. Kurucz in 2012 based on 
the new observed data of P~{\sc ii} levels, by which the previous Kurucz \& Bell's 
(1995) value of +0.384 was somewhat revised. 
This VALD value is also quite consistent with the data (+0.42) of Wiese et al. (1969),
which is also included in the database of NIST (National Institute of Standards 
and Technology). 
Therefore, it is unlikely that our $A^{\rm N}$ results are significantly
underestimated due to the use of an erroneous $gf$ value.

Another possibility is that the actual solar P abundance might be lower 
than the currently believed value ($A_{\odot} \sim$~5.4--5.5).
This problem is separately focused upon in Appendix~A.
As discussed therein (cf. Sect.~A5), although a possibility can not be ruled 
out that $A_{\odot}$ could be somewhat reduced by $\lesssim$~0.1--0.2~dex 
(i.e., down to $\sim 5.3$), since uncertainties are still involved with 
P~{\sc i} line formation calculations (e.g., H~{\sc i} collision rates, 
treatment of 3D effect), there is no convincing reason for such a downward 
revision of $A_{\odot}$.
Moreover, if the solar photospheric P abundance were to be appreciably changed,
another problem of discrepant $A_{\odot}$ from $A_{\rm meteorite}$ would 
newly emerge, because a good agreement has already been accomplished between 
these two (Asplund et al. 2009).

Therefore, frankly accepting the result of the analysis, we conclude that 
the phosphorus abundances of B-type stars ($\langle A^{\rm N} \rangle$) are 
systematically lower than that of the Sun ($A_{\odot}$) by $\sim$~0.2--0.3~dex.
The cause for such a puzzling discrepancy (apparently contradicting 
the scenario of galactic chemical evolution) would be worth further earnest 
investigation. Meanwhile, follow-up studies by other researchers are also 
desirably awaited for an independent check of this observational finding.
 
\section{Summary and Conclusion}

Recently, special attention is being focused on the abundance of phosphorus
in the universe, mainly because of its astrobiological importance as a key 
element for life. For example, whether or not a sufficient amount of P 
(required for the rise of life) remains on the surface of a planet depends 
critically upon the primordial P abundance of the material, from which a star 
and the associated planetary system were formed.

Stimulated by such increasing interest, a number of spectroscopic studies 
intending to establish stellar P abundances have been published lately.
These P-specific investigations are directed to late-type (FGK-type) stars of 
lower mass (around $\sim 1 M_{\odot}$), which are generally long-lived and 
thus retain the information of P composition in comparatively earlier time 
of the Galaxy ($\sim 10^{9}$--$10^{10}$~yr ago) when they were formed.

However, little efforts have been made so far to comprehensive P abundance
determinations of young hotter stars, such as B-type stars of 
$\sim$~3--10~$M_{\odot}$,  which reflect the gas composition of the Galaxy 
in the more recent past (several times $\sim 10^{7}$--$10^{8}$~yr ago). 
This is presumably because P lines of sufficient strength usable as abundance
indicators are scarce. Actually, many of those B-type stars for which 
such determinations were done are P-rich chemically peculiar stars (HgMn 
stars), while the number of normal B-type stars with known P abundances   
is quite limited. 

Thus, motivated by the necessity of clarifying the behavior of phosphorus 
in hotter stars of higher mass, P abundances of $\sim 80$ apparently 
bright sharp-lined early-to-late B-type stars on the upper main sequence 
were determined from the P~{\sc ii} 6043.084~\AA\ line (the strongest 
P~{\sc ii} line in the optical region), with an aim of getting information
on the composition of this element in the young galactic gas in comparison 
with the abundance of the older Sun (age of 4.6$\times 10^{9}$~yr).

A special emphasis was placed upon taking into account the non-LTE effect
based on extensive statistical-equilibrium calculations on P~{\sc ii} atoms, 
since the assumption of LTE was assumed in all the previous P abundance 
determinations.

Regarding the procedures of analysis, a spectrum-fitting was first applied 
to the wavelength region comprising the P~{\sc ii} 6043 line, and its 
equivalent width ($W_{6043}$) was then derived from the fitting-based 
abundance solution for each star. Finally, non-LTE abundance/correction 
as well as possible error were evaluated from such established $W_{6043}$.
An inspection of the resulting P abundances revealed that the program stars 
are divided into two groups.

The first group is those showing a considerable overabundance of P 
(supersolar by $\sim$~0.5--1.5~dex), the extent of which progressively 
increases with $T_{\rm eff}$. These P-rich stars are observed 
at $T_{\rm eff} \lesssim 16000$~K (late B-type) and mostly belong to 
chemically peculiar stars of HgMn-type.

The second group consists of normal B-type stars, whose P abundances are 
comparatively homogeneous without such a prominent P anomaly as in the first 
group. However, different trends are observed between the LTE and non-LTE cases. 
(i) Though the LTE abundances tend to distribute around the solar value, 
they show a slight gradient (i.e., increasing with a decrease in $T_{\rm eff}$ 
at $T_{\rm eff} \gtrsim 16000$~K). 
(ii) Meanwhile, this systematic trend disappears in the non-LTE 
abundances which are satisfactorily uniform, because of the cancellation 
due to the $T_{\rm eff}$-dependent negative non-LTE corrections (amounting 
to $\sim$~0.1--0.5~dex). 

This $T_{\rm eff}$-independent nature seen in the non-LTE abundances 
of normal B stars suggests that they represent the P composition of 
the galactic gas at the time when these young stars were born (some 
$\sim 10^{7}$--$10^{8}$~yr ago).
One puzzling problem is, however, that these non-LTE abundances (around $\sim 5.2$) 
are appreciably lower than the P abundance of the Sun (formed $\sim 4.6 \times 10^{9}$~yr 
ago) by $\sim$~0.2--0.3~dex, which means that the galactic gas composition of P  
has decreased with time in contradiction to the concept of chemical evolution.

Although other possibilities (e.g., error in the adopted $gf$ value of 
P~{\sc ii} 6043 line?, inadequacy in the current solar P abundance?) 
were also examined, they do not seem to be so likely.
It may thus be concluded that the discrepancy of P abundance between the Sun 
and B-type stars really exists, the cause of which should be further investigated. 

\Acknow{This investigation has made use of the SIMBAD database, operated by CDS, 
Strasbourg, France, and the VALD database operated at Uppsala University,
the Institute of Astronomy RAS in Moscow, and the University of Vienna.}

\newpage

\appendix{{\bf Appendix A: On the Solar Photospheric Abundance of Phosphorus}}\\

\subsection*{A1. Literature Values of Solar P Abundance}

Regarding the phosphorus abundance in the solar photosphere (usually
derived from P~{\sc i} lines in the $Y$- or $H$-band of the near-IR region),
not a few spectroscopic studies have been published over the past half century, 
in which rather similar $A_{\odot}$ values of $\simeq$~5.4--5.5 are reported.
See Table~2 of Caffau et al. (2007) for a summary of 8 values (5.43, 5.45, 5.45,
5.45, 5.49, 5.45, 5.36, and 5.46) published before 2007. Thereafter, 
Asplund et al. (2009) presented a revised value of $A_{\odot} = 5.41$.
The Anders \& Grevesse's (1989) value of $A_{\odot} = 5.45$ adopted in 
this paper (cf. footnote~3) is almost the same as the mean of these values.

However, all these determinations were done with the assumption of LTE, 
given the lack of information regarding the non-LTE effect on the P~{\sc i} 
lines  so far. Therefore, non-LTE calculations for neutral phosphorus 
were newly carried out in order to elucidate how and whether the non-LTE
corrections are important in P abundance determinations for the Sun.
In addition, how this effect would depend upon the atmospheric parameters
is also briefly examined in scope of application to late-type stars in general. 

\subsection*{A2. Atomic Model}

The adopted model atom of P~{\sc i} comprises 56 terms (up to 3s$^{2}$\,3p$^{2}$\,5d\,$^{4}$D
at 79864 cm$^{-1}$) and 761 radiative transitions, which was constructed in a similar 
manner to the case of P~{\sc ii} described in Sect.~5. The contribution of P~{\sc ii} 
was taken into account in the number conservation of total P atoms. 

Regarding the photoionization cross section, Tayal's (2004) theoretically calculated 
data were adopted for the lowest three terms ($^{4}$S$^{\circ}$, $^{2}$D$^{\circ}$, 
$^{2}$P$^{\circ}$; read from Fig.~1--3, Fig.~9, and Fig.~10 of his paper), while the 
hydrogenic approximation was assumed for the other terms.

As to the collisional rates, the recipe described in Sect.~3.1.3 of Takeda (1991)
was basically followed. The collision rates due to neutral hydrogen atoms 
(which are important in late-type stars but subject to large uncertainties) were 
computed by Steenbock \& Holweger's (1984) formula (based on the classical Drawin's 
cross section), which can be further multiplied by a correction factor ($k$) 
if necessary. Although we adopt $k=1$ (use of classical formula unchanged)
as the standard choice, a special case of $k=10^{-3}$ (considerably reduced 
to a negligible level) was also tried in order to see its importance.

\subsection*{A3. Non-LTE Effect on P~I 10581 Line in FGK-type Stars}

First, the calculations were done for 10 solar-metallicity models (ATLAS9 models 
by Kurucz 1993) resulting from combinations of ($T_{\rm eff}$ =  4500, 5000, 5500, 
6000, 6500~K) and ($\log g$ = 2.0, 4.0), while assuming $\xi = 2$~km~s$^{-1}$
and [P/H] = 0 ($A = 5.45$).
The runs of $l_{0}^{\rm NLTE}/l_{0}^{\rm LTE}$ and $S_{\rm L}/B$ with depth
for the transition corresponding to P~{\sc i} 10581.58 line (representative 
P~{\sc i} line in the near-IR region) are shown in Fig.~8, while Fig.~9 displays 
how $W_{10581}$ (equivalent widths calculated in LTE and non-LTE; upper panel (a)) 
as well as $\Delta_{10581}$ (non-LTE correction; lower panel (b))
depend upon $T_{\rm eff}$ and $\log g$.
The following characteristics are read from these figures.
\begin{itemize}
\item
The line is generally intensified by the non-LTE effect ($W^{\rm N} > W^{\rm L}$ 
and $\Delta < 0$; cf. Fig.~9), because both $l_{0}^{\rm NLTE}/l_{0}^{\rm LTE} > 1$ 
and $S_{\rm L}/B < 1$ (cf. Fig.~8) act in the direction of line strengthening. 
\item
This non-LTE effect becomes progressively larger with an increase in $T_{\rm eff}$ 
and with a decrease in $\log g$ (Fig.~9b).
\item
How the neutral hydrogen collision is treated has an appreciable impact on the non-LTE 
correction. If the standard value of H~{\sc i} collision is practically neglected by 
a drastic reduction ($k=10^{-3}$), $|\Delta|$ increases by $\sim$~0.1--0.2~dex (Fig.~9b). 
\item
In summary, non-LTE corrections had better be taken into account in P abundance 
determinations from P~{\sc i} lines for FGK stars, especially for those of 
comparatively higher $T_{\rm eff}$ or lower $\log g$, for which significant 
negative corrections amounting to several tenths dex may be expected. 
\end{itemize}

%Fig. 8
\begin{figure}[h]
\begin{minipage}{150mm}
\begin{center}
\includegraphics[width=9.0cm]{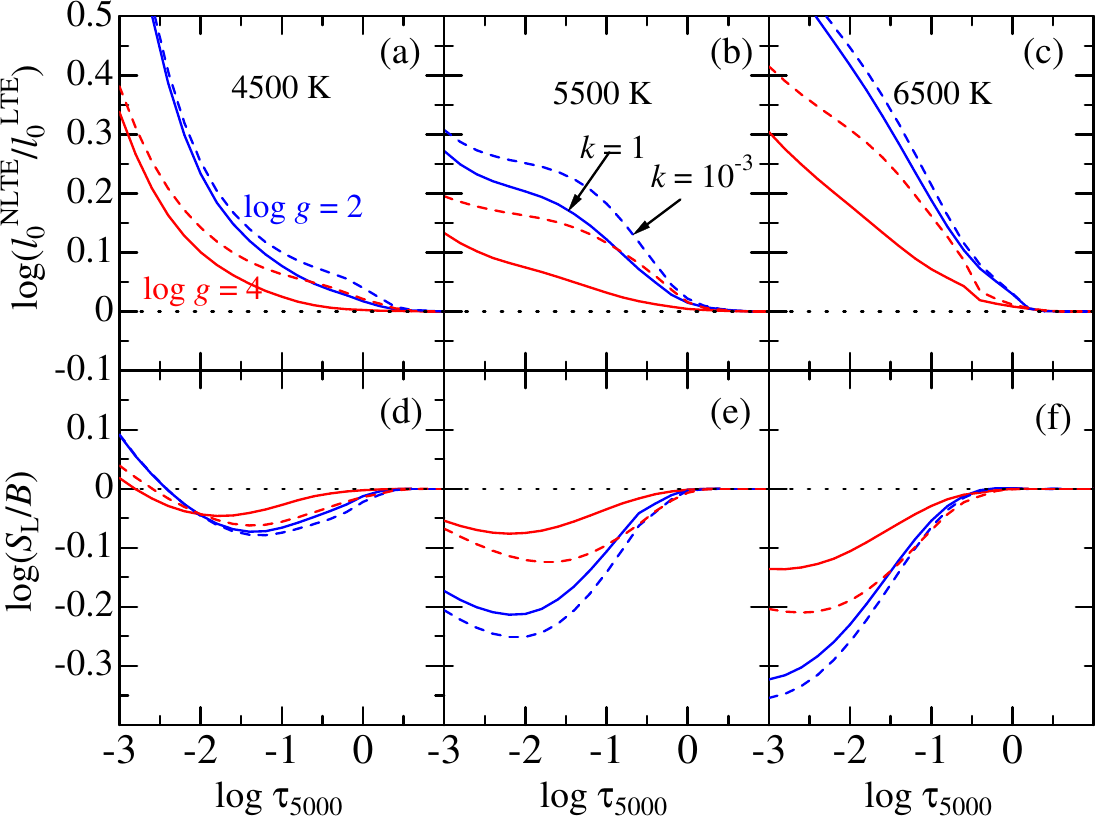}
\end{center}
\FigCap{
The non-LTE-to-LTE line-center opacity ratio (upper panels (a)--(c)) and 
the ratio of the line source function ($S_{\rm L}$) 
to the local Planck function ($B$) (lower panels (d)--(f))  
for the P~{\sc i} 4s\,$^{4}$P$^{\circ}$\,--\,4s\,$^{4}$D transition 
(corresponding to P~{\sc i} 10581 line) of multiplet~1, 
plotted against the continuum optical depth at 5000~\AA. 
Shown here are the calculations done with $\xi = 2$~km~s$^{-1}$ 
and the solar P abundance ([P/H] = 0) on the solar-metallicity models 
of $T_{\rm eff}$ = 4500~K (left panels (a), (d)), 
5500~K (middle panels (b), (e)), and 6500~K (right panels (c), (f)). 
for two surface gravities of $\log g$ = 2 and 4.
At each panel, the results for different treatment of neutral hydrogen
collisions ($k = 1$ and $10^{-3}$) are discriminated by line-types 
(solid and dashed lines), while those of $\log g = 2.0$ and 4.0 are 
depicted by different colors (blue and red, respectively). 
}
\end{minipage}
\end{figure}

%Fig. 9
\begin{figure}[h]
\begin{minipage}{150mm}
\begin{center}
\includegraphics[width=8.0cm]{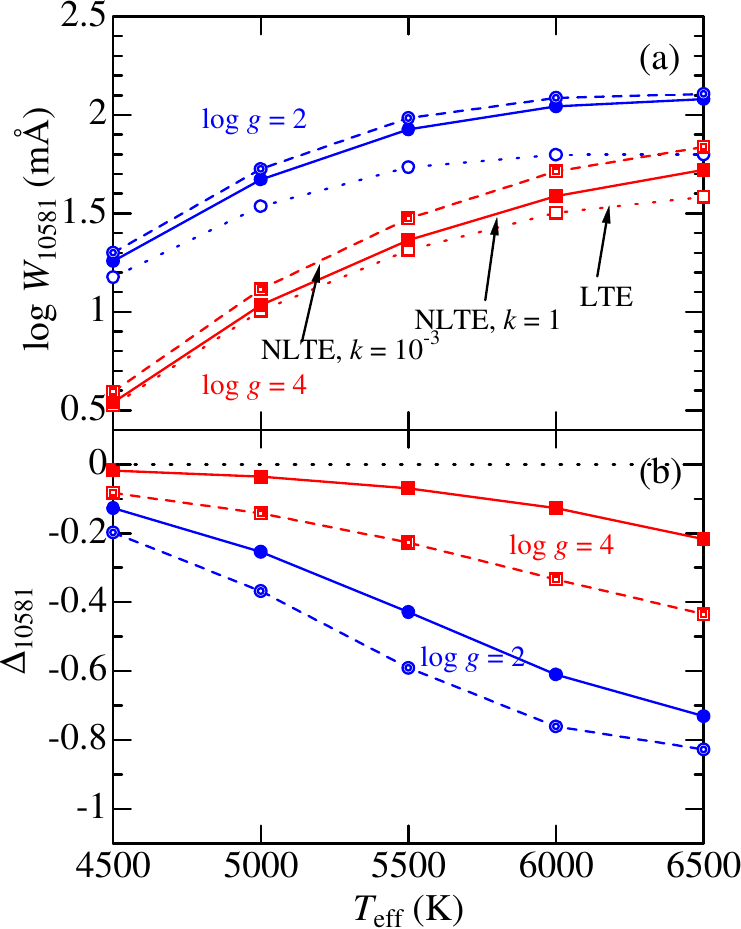}
\end{center}
\FigCap{
Theoretical equivalent widths ($W$) calculated for the P~{\sc i} 10581 
line and the corresponding non-LTE corrections are plotted
against $T_{\rm eff}$. In panel (a), the runs of $W^{\rm L}$ (LTE),
$W_{k=1}^{\rm N}$ (non-LTE with $k = 1$), and 
$W_{k=10^{-3}}^{\rm N}$ (non-LTE with $k = 10^{-3}$) are depicted
in dotted, solid, and dashed lines, respectively.
Similarly, the runs of $\Delta_{k=1}$ (solid lines) and 
$\Delta_{k=10^{-3}}$ (dashed lines) are shown in panel (b). 
These calculations were done with the solar P abundance
([P/H] = 0.0) on 10 solar metallicity models for two $\log g$ 
(2 and 4) and five $T_{\rm eff}$ (4500, 5000, 5500, 6000, 
and 6500~K) values. The results for $\log g = 2$ and 4 are 
depicted in blue and red, respectively. 
}
\end{minipage}
\end{figure}

\subsection*{A4. Reanalysis of Solar P~I Lines}

Then, statistical equilibrium calculation was done for Kurucz's (1993) ATLAS9 solar 
model atmosphere ($T_{\rm eff}$ = 5780~K, $\log g = 4.44$, solar metallicity) with [P/H] = 0.0,
in order to reanalyze the solar P~{\sc i} lines by taking into account the non-LTE effect. 
The basic data for the solar equivalent widths and $\log gf$ values of 15 P~{\sc i}
lines were taken from Table~3 of Bi\'{e}mont et al. (1994), which are the disk-center
values measured from Jungfraujoch Atlas ($W_{\lambda}^{\rm d.c.}$) and transition 
probabilities based on their refined calculations. Regarding the solar microturbulence,
a reasonable value of $\xi_{\odot} = 1$~km~s$^{-1}$ was adopted as often assumed.
The resulting P abundances and non-LTE corrections are summarized in Table~2,
from which the following conclusions can be drawn.

%Table~2
\setcounter{table}{1}
\begin{table}[h]
\caption{Non-LTE analysis of solar photospheric P abundances.}
%\small
\scriptsize
\begin{center}
\begin{tabular}{cccccrccccc}\hline\hline
Mult. & Transition & $^{*}\lambda$ & $^{*}\chi_{\rm low}$ & $^{\dagger}\log gf$ & $^{\dagger}W^{\rm d.c.}_{\lambda}$ &
$A^{\rm L}$ & $A^{\rm N}_{k=1}$ & $\Delta_{k=1}$ & $A^{\rm N}_{k=10^{-3}}$ & $\Delta_{k=10^{-3}}$ \\
(1) & (2) & (3) & (4) & (5) & (6) & (7) & (8) & (9) & (10) & (11) \\
\hline
   3 & 4s\,$^{4}{\rm P}_{5/2}$\,--\,4p\,$^{4}{\rm S}_{3/2}$ &   9525.741 &  6.985 & $-0.100$ &   7.7 &  5.431 &  5.394 & $-0.037$ &  5.297 & $-0.134$ \\
   2 & 4s\,$^{4}{\rm P}_{3/2}$\,--\,4p\,$^{4}{\rm P}_{1/2}$ &   9750.748 &  6.954 & $-0.180$ &   6.3 &  5.421 &  5.388 & $-0.033$ &  5.296 & $-0.125$ \\
   4 & 4s\,$^{2}{\rm P}_{1/2}$\,--\,4p\,$^{2}{\rm P}_{3/2}$ &   9790.194 &  7.176 & $-0.690$ &   0.9 &  5.245 &  5.223 & $-0.022$ &  5.149 & $-0.096$ \\
   2 & 4s\,$^{4}{\rm P}_{5/2}$\,--\,4p\,$^{4}{\rm P}_{5/2}$ &   9796.828 &  6.985 & $+0.270$ &  16.9 &  5.514 &  5.475 & $-0.039$ &  5.372 & $-0.142$ \\
   4 & 4s\,$^{2}{\rm P}_{1/2}$\,--\,4p\,$^{2}{\rm P}_{1/2}$ &   9903.671 &  7.176 & $-0.300$ &   3.8 &  5.583 &  5.559 & $-0.024$ &  5.482 & $-0.101$ \\
   2 & 4s\,$^{4}{\rm P}_{5/2}$\,--\,4p\,$^{4}{\rm P}_{3/2}$ &   9976.681 &  6.985 & $-0.290$ &   2.9 &  5.604 &  5.573 & $-0.031$ &  5.485 & $-0.119$ \\
   4 & 4s\,$^{2}{\rm P}_{3/2}$\,--\,4p\,$^{2}{\rm P}_{1/2}$ &  10204.716 &  7.213 & $-0.520$ &   1.9 &  5.468 &  5.444 & $-0.024$ &  5.369 & $-0.099$ \\
   1 & 4s\,$^{4}{\rm P}_{1/2}$\,--\,4p\,$^{4}{\rm D}_{3/2}$ &  10511.588 &  6.936 & $-0.130$ &   7.8 &  5.381 &  5.348 & $-0.033$ &  5.252 & $-0.129$ \\
   1 & 4s\,$^{4}{\rm P}_{3/2}$\,--\,4p\,$^{4}{\rm D}_{5/2}$ &  10529.524 &  6.954 & $+0.240$ &  15.3 &  5.365 &  5.329 & $-0.036$ &  5.224 & $-0.141$ \\
   1 & 4s\,$^{4}{\rm P}_{5/2}$\,--\,4p\,$^{4}{\rm D}_{7/2}$ &  10581.577 &  6.985 & $+0.450$ &  24.7 &  5.436 &  5.395 & $-0.041$ &  5.279 & $-0.157$ \\
   1 & 4s\,$^{4}{\rm P}_{1/2}$\,--\,4p\,$^{4}{\rm D}_{1/2}$ &  10596.903 &  6.936 & $-0.210$ &  10.4 &  5.586 &  5.552 & $-0.034$ &  5.453 & $-0.133$ \\
   1 & 4s\,$^{4}{\rm P}_{3/2}$\,--\,4p\,$^{4}{\rm D}_{3/2}$ &  10681.406 &  6.954 & $-0.190$ &   7.8 &  5.434 &  5.401 & $-0.033$ &  5.305 & $-0.129$ \\
   1 & 4s\,$^{4}{\rm P}_{3/2}$\,--\,4p\,$^{4}{\rm D}_{1/2}$ &  10769.511 &  6.954 & $-1.070$ &   1.0 &  5.358 &  5.329 & $-0.029$ &  5.240 & $-0.118$ \\
   1 & 4s\,$^{4}{\rm P}_{5/2}$\,--\,4p\,$^{4}{\rm D}_{5/2}$ &  10813.141 &  6.985 & $-0.410$ &   4.0 &  5.338 &  5.308 & $-0.030$ &  5.216 & $-0.122$ \\
---  & 4s\,$^{2}{\rm P}_{3/2}$\,--\,4p\,$^{2}{\rm D}_{5/2}$ &  11183.240 &  7.213 & $+0.400$ &  10.6 &  5.272 &  5.246 & $-0.026$ &  5.170 & $-0.102$ \\
\hline
\end{tabular}
\end{center}
(1) Multiplet number (Moore 1959). (2) Spectroscopic designation of the lower and upper levels 
of the transition. (3)  Air wavelengths (in \AA). (4) Lower excitation potential (in eV).
(5) Logarithm of the statistical weight of the lower level times the oscillator strength.
(6) Solar equivalent widths at the disk center (in m\AA). (7) LTE abundance (in dex).
(8) Non-LTE abundance for $k = 1$ (in dex). (9) Non-LTE correction for $k = 1$ (in dex).
(10)  Non-LTE abundance for $k = 10^{-3}$ (in dex). (11) Non-LTE correction for $k = 10^{-3}$ (in dex).\\
$^{*}$ Taken from VALD (Ryabchikova et al. 2015).\\
$^{\dagger}$ Taken from Bi\'{e}mont et al. (1994).
\end{table}
The extents of (negative) non-LTE corrections are only a few hundredths dex ($k=1$ case) 
and thus not particularly important, though $|\Delta|$ may somewhat increase further by 
$\sim 0.1$ if H~{\sc i} collision is neglected ($k = 10^{-3}$). This is mainly due to the 
high-gravity nature of the Sun ($\log g = 4.44$) as a dwarf, since $|\Delta|$ appreciably 
decreases with an increase in $\log g$ (cf. Fig.~9b).\footnote{Actually, $|\Delta|$  
in this case (disk-center intensity spectrum) is somewhat smaller than that obtained 
by an interpolation/extrapolation of Fig.~9b (calculated for the flux spectrum),
because the non-LTE effect becomes less significant for the former case of 
deeper-forming lines than the latter.} 
Therefore, the speculation addressed by Asplund et al. (2009) ``departures from LTE 
are not expected to be significant for P'' (based on an analogy with the S~{\sc i} 
case) may be regarded as reasonable.  

The mean P abundances averaged over 15 lines are $\langle A^{\rm L} \rangle = 5.43$ 
(LTE abundance), $\langle A^{\rm N}_{k=1} \rangle = 5.40$ (non-LTE abundance for $k=1$),
and  $\langle A^{\rm N}_{k=10^{-3}} \rangle = 5.31$ (non-LTE abundance for $k=10^{-3}$),
where the standard deviation is $\sigma = 0.11$ for all the three cases.
This $\langle A^{\rm L} \rangle$ (5.43) is reasonably consistent with Bi\'{e}mont et al.'s 
(1994) result of 5.45 (obtained by using the same lines with the same $W_{\lambda}^{\rm d.c.}$ 
and $\log gf$). According to the standard non-LTE abundance of 5.40 
($\langle A^{\rm N}_{k=1} \rangle$) obtained here, we may state that the previously
reported results of solar P abundance (cf. Sect.~A1) can not be significantly 
revised even when the non-LTE correction (a few hundredths dex) is taken into account.

\subsection*{A5. Is Downward Revision of $A_{\odot}$ Possible?}

In view of discrepancy between the P abundances of normal B-type stars and that of 
the Sun (the former being systematically lower than the latter by $\sim$~0.2--0.3~dex) 
discussed in Sect.~7.2, some discussion about whether the actual $A_{\odot}$ 
could be lower than the currently accepted value may be in order. 
  
Neglecting the H~{\sc i} collision would further reduce $A^{\rm N}$ by $\sim 0.1$~dex 
down to $\sim 5.3$. However, there is no justification for such 
a drastic reduction of the classical rates for rather high-excitation P~{\sc i} lines
under question.\footnote{Admittedly, such a situation does exist for the case of other lines. 
For example, it is known that neutral-hydrogen collision rates calculated by using the 
classical cross section are considerably overestimated for the resonance lines of 
alkali elements (e.g., Li~{\sc i} 6708 or K~{\sc i} 7699).} Therefore, much can not be
said about this possibility until more information of H~{\sc i} collision 
rates for P~{\sc i} is obtained (preferably based on up-to-date quantum-mechanical 
calculations).

Alternatively, it was once considered that inclusion of the 3D effect 
might reduce $A_{\odot}$ by $\lesssim 0.1$~dex, since Asplund et al. (2005) derived an 
appreciably lower value of $A_{\odot} = 5.36$ by including the 3D-effect. 
However, such a low-scale result was not confirmed by Caffau et al.'s (2007) 
new 3D analysis which resulted in $A_{\odot} = 5.46 (\pm 0.04)$, showing that 
the 3D correction is insignificant (only a few hundredths dex) for the solar 
P abundance determination. Thus, this possibility is not very prospective, either.

In any case, although a possibility of solar P abundance being reduced by 
$\lesssim$~0.1--0.2 (i.e., down to $\sim 5.3$) can not be excluded, 
the solar P abundance would then be discrepant from that of meteorite.
That is, since the old $A_{\rm meteorite}$(P) value of $5.57 \pm 0.04$ derived by 
Anders \& Grevesse (1989) was revised by Asplund et al. (2009) as 
$5.43 \pm 0.04$ (based on the data of CI carbonaceous chondrites taken from 
Lodders et al. 2009), a good agreement between $A_{\rm meteorite}$ and 
$A_{\odot}$ is now accomplished. This consistency would break down
if $A_{\odot}$ were appreciably changed.  


\vspace*{10mm}

\begin{references}
\refitem{Alecian, G., \& Stift, M. J.}{2019}{MNRAS}{482}{4519}\\
\hspace*{-10mm}Allen, C. S., 1998, {\it Abundance analysis of normal and 
mercury-manganese type late-B stars from optical spectra} (PhD Thesis: 
University College London) [https://discovery.ucl.ac.uk/id/eprint/10097529/].
\refitem{Anders, E., \& Grevesse, N.}{1989}{Geochim. Cosmochim. Acta}{53}{197}\\
\hspace*{-10mm}Asplund, M., Grevesse, N., \& Sauval, A. J., 2005, 
in {it Cosmic Abundances as Records of Stellar Evolution and Nucleosynthesis,
ASP Conf. Ser.}, {\bf 336}, 25.
\refitem{Asplund, M., Grevesse, N., Sauval, A. J., \& Scott, P.}{2009}{ARA\&A}{47}{481}
\refitem{Bekki, K., \& Tsujimoto, T.}{2024}{ApJL}{967}{L1}
\refitem{Bi\'{e}mont, E., Martin, F., Quinet, P., \& Zeippen, C. J.}{1994}{A\&A}{283}{339}
\refitem{Caffau, E., Steffen, M., Sbordone, L., Ludwig, H.-G., \& Bonifacio, P.}
  {2007}{A\&A}{473}{L9}
\refitem{Catanzaro, G., Giarrusso, M., Leone, F., Munari, M., Scalia, C.,
  Sparacello, E., \& Scuderi, S.}{2016}{MNRAS}{460}{1999}\\
\hspace*{-10mm}Cayrel, R. 1988, in {\it The Impact of Very High S/N Spectroscopy on Stellar Physics, 
Proceedings of IAU Symposium 132}, ed. G. Cayrel de Strobel, M. Spite (Kluwer, Dordrecht), p.345.
\refitem{Fossati, L., Ryabchikova, T., Bagnulo, S., Alecian, E., Grunhut, J.,
  Kochukhov, O., \& Wade, G.}{2009}{A\&A}{503}{945}
\refitem{Ghazaryan, S., \& Alecian, G.}{2016}{MNRAS}{460}{1922}
\refitem{Golriz, S. S., \& Landstreet, J. D.}{2017}{MNRAS}{466}{1597}
\refitem{Hinkel, N. R., Hartnett, H. E., \& Young, P.}{2020}{ApJL}{900}{L38}\\
\hspace*{-10mm}Hoffleit, D. \& Jaschek, C., 1991, {\it The Bright Star Catalogue, 
5th revised edition}, (New Haven, Conn.: Yale University Observatory).\\
\hspace*{-10mm}Kurucz, R. L. 1993, {\it Kurucz CD-ROM}, No. 13  
(Harvard-Smithsonian Center for Astrophysics).\\
\hspace*{-10mm}Kurucz, R. L., \& Bell, B. 1995, {\it Kurucz CD-ROM}, No. 23  
(Harvard-Smithsonian Center for Astrophysics).
\refitem{Lejeune, T., \& Schaerer, D.}{2001}{A\&A}{366}{538}\\
\hspace*{-10mm}Lodders, K., Palme, H., \& Gail, H.-P. 2009,
{\it Solar System, Landolt-B\"{o}rnstein - Group VI Astronomy and Astrophysics, 
Volume 4B}, p.~712 (Springer, Berlin).
\refitem{Maas, Z. G., Hawkins, K., Hinkel, N. R., Cargile, P., 
  Janowiecki, S., \& Nelson, T.}{2022}{AJ}{164}{61}\\
\hspace*{-10mm}Moore, C. E.. 1959, {\it A multiplet Table of 
Astrophysical Interest: NBS Technical Note No. 36, Reprinted Version
of the 1945 edition} (U. S. Department of Commerce, Washington).
\refitem{Nahar, S. N., Hern\'{a}ndez, E. M., Hern\'{a}ndez, L., et al.}{2017}{JQSRT}{187}{215}
\refitem{Ndiaye, M. L., LeBlanc, F., \& Khalack, V.}{2018}{MNRAS}{477}{3390}
\refitem{Niemczura, E., Morel, T., \& Aerts, C.}{2009}{A\&A}{506}{213}
\refitem{Peters, G. J., \& Aller, L. H.}{1970}{ApJ}{159}{525}\\
\hspace*{-10mm}Peters, G. J., \& Polidan, R. S., 1985, {\it
Proc. IAU Symp. 111, Calibration of Fundamental Stellar Quantities}
(Reidel, Dordrecht), p.~417.
\refitem{Pintado, O. I., \& Adelman, S. J.}{1993}{MNRAS}{264}{63}
\refitem{Przybilla, N., Butler, K., Becker, S. R., \& Kudritzki, R. P.}{2006}{A\&A}{445}{1099}
\refitem{Ryabchikova, T., Piskunov, N., Kurucz, R.~L., Stempels, H.~C., Heiter, U., 
  Pakhomov, Yu, \& Barklem, P.~S.}{2015}{Phys. Scr.}{90}{054005}
\refitem{Sadakane, K., \& Nishimura, M.}{2022}{PASJ}{74}{298}
\refitem{Steenbock, W., \& Holweger, H.}{1984}{A\&A}{130}{319}
\refitem{Takeda, Y.}{1991}{A\&A}{242}{455}
\refitem{Takeda, Y.}{2023}{Acta Astron.}{73}{35}
\refitem{Takeda, Y., Kambe, E., Sadakane, K., \& Masuda, S.}{2010}{PASJ}{62}{1239 (Paper I)}
\refitem{Takeda, Y., Kawanomoto, S., \& Ohishi, N.}{2014}{PASJ}{66}{23 (Paper II)}
\refitem{Tayal, S. S.}{2004}{J. Phys. B; At. Mol. Opt. Phys.}{37}{3593}\\
\hspace*{-10mm}Wiese, W. L., Smith, M. W., \& Miles, B. M. 1969, 
{\it Atomic transition probabilities, Vol. II: Sodium through calcium - 
A critical data compilation, in Nat. Stand. Ref. Data Ser., NSRDS-NBS 22}
(U.S. Government Printing Office, Washington, D.C.).
\end{references}
\end{document}